\documentclass[%
twocolumn,
amsmath,amssymb,
aps,
]{revtex4-2}

\usepackage{graphicx}% Include figure files
\usepackage{dcolumn}% Align table columns on decimal point
\usepackage{bm}% bold math
\usepackage{hyperref}% add hypertext capabilities
\usepackage[dvipsnames]{xcolor}
\usepackage{comment}
\usepackage{multirow}

\def\beq{\begin{equation}}
\def\eeq{\end{equation}}
\def\bsp#1\esp{\begin{split}#1\end{split}}
\def\bal#1\eal{\begin{align}#1\end{align}}

\newcommand\rd   {\ensuremath{\mathrm{d}}}
\newcommand\re   {\ensuremath{\mathrm{e}}}

\begin{document}

\title{
Derivative expansion for
computing critical exponents
of $O(N)$ symmetric models
at NNLO 
}

\author{Zolt\'an P\'eli}%
 \email{zoltanpeli92@gmail.com}
\affiliation{%
 MTA-DE Particle Physics Research Group,\\ H-4010 Debrecen, PO Box 105, Hungary
}%

\date{\today}

\begin{abstract}
We apply the derivative expansion of the effective action in the exact renormalization group equation up to fourth order 
to the $Z_2$  and $O(N)$ symmetric scalar models in $d=3$ Euclidean dimensions.
We compute the critical exponents $\nu$, $\eta$ and $\omega$ using polynomial expansion in the field. 
We obtain our predictions for the exponents employing two regulators widely used in ERG computations. We apply Wynn's epsilon algorithm to improve the predictions for the critical exponents, extrapolating beyond the next-to-next-to-leading order prediction of the derivative expansion.
\end{abstract}

\maketitle

\section{Introduction}
In this work we compute the critical exponents $\nu$, $\eta$ and $\omega$ for the $Z_2$ and $O(N)$ symmetric scalar models in $d=3$ Euclidean dimensions. We use the exact renormalization group (ERG) equation for effective average action \cite{WETTERICH199390}. The exact renormalization group (ERG) is a highly versatile method for tackling problems in statistical physics and quantum field theory. Its modern formulation has sprouted from Wilson's approach to renormalization \cite{WILSON197475}.

There are a number of other ways in modern physics to obtain critical exponents. Perhaps the first one to come to mind is lattice simulation. The Monte-Carlo (MC) simulations provide one of the most precise determination of the exponents for the Ising \cite{MC-ising} and XY \cite{MC-XY} universality classes. Generally, a larger lattice yields more precise predictions, but also increases the computational effort. The most commonly applied method in quantum field theory is the loop-expansion, which requires a smallness of the couplings in the Lagrangian. In fixed $d=3$ dimensions, the Ising exponents have been computed up to six-loop order \cite{Guida_1998} and the beta functions are determined at seven loops \cite{Loop-ising}. Wilson's $d=4-\epsilon$ expansion has also been applied up to $\epsilon^6$ \cite{eps-expand}. 
Presently, the most precise computation for the Ising exponents comes from the conformal bootstrap method (CB) \cite{bootstrap-ising} using conformal field theory. This method also has a high computational cost, see Tab. II. of \cite{CB-cost} for instance. The last highlight on this list is the large-$N$ expansion.
It is applicable on theories, where the symmetry group corresponding to the symmetry of the Lagrangian is $O(N),~SO(N),~SU(N)$ and so on. 

The ERG is formulated in terms of functional equations, which are in general very hard to solve. In order to tackle this difficulty, a precise approximation scheme has to be applied, which is most often the so called derivative expansion (DE). The DE consists of expanding the action in terms of the gradient of the field. This approximation scheme contains no explicit small parameter, thus its convergence has been questioned. 
Recently however,
arguments have been put forward that the DE is indeed convergent \cite{PhysRevLett.123.240604} at least for the $Z_2$ and $O(N)$ symmetric models. The corrections were shown to be dampened by a factor of $1/4 \sim 1/9$, depending on the regulator function. The physical predictions depend on the regulator function at fixed order in the DE. 
This is similar to the renormalization scale dependence in perturbative quantum field theory. 

Here we compute the critical exponents at the next-to-next-to-leading order (NNLO) of the DE on the $Z_2$ symmetric scalar model as a benchmark and then generalize the computations to the $O(N)$ symmetric models.
Our results complement those of Ref.~\cite{full-on}, where the authors employ the DE at NNLO as well, but there are key differences: (i) we do not use truncation of momenta in the derivation of our beta-functions, (ii) we employ Taylor expansion of the beta-functions in the field. These beta-functions describe the scale dependence of different functions depending on the field. The Taylor expansion reduces these to the beta-functions for coupling strengths corresponding to different vertices of the field. We compute the exponents with the exponential regulator, which is applicable at any order of the derivative expansion and (iii) also with a $\Theta$-type regulator \cite{Litim2001}, which is the simplest applicable regulator at NNLO. 
The critical exponents of the $Z_2$ symmetric model have already been computed in Ref.~\cite{dela2003} using Taylor expansion in the field, although with a more severe truncation of the Taylor series.

By increasing the number of terms in the Taylor expansion of the scale dependent functions, the values of the critical exponents fluctuate and eventually stabilize around their limiting values. 
Reassuringly, similar behaviour has been observed in Ref.~\cite{dela2003}.
Interestingly, we find that the exponents $\nu$, $\eta$ and $\omega$ of the $O(N)$ symmetric model are estimated remarkably well even at the zeroth order of the Taylor expansion in the field variable of the scale dependent functions corresponding to the NNLO of the DE. Furthermore, this fluctuation of the exponents at the NNLO is much less pronounced in the $O(N)$ symmetric case than in the $Z_2$ symmetric one.
This dampening of the fluctuation is likely the result of having more scale dependent functions for the $O(N)$ symmetric models than for the $Z_2$ symmetric one. These scale dependent functions have to interplay in such a way, that the predictions for the exponents are in good agreement with other method's predictions. This is true for at least large values of $N$, where the higher order contributions from the DE are expected to be very small as the leading-order of the DE becomes exact in the limit $N\to \infty$ \cite{largeN-refree}.

We introduce the ERG briefly in Sec.~\ref{sec:erg}. The procedure we use to acquire the results is outlined in Sec.~\ref{sec:beta}. Our findings for the $Z_2$ symmetric model are detailed in Sec.~\ref{sec:z2}, while those of the $O(N)$ symmetric one can be found in Sec.~\ref{sec:on}. 

\section{Exact Renormalization Group
}
\label{sec:erg}

The ERG uses functional integro-differential equations to describe the dependence of a theory on the variation of the characteristic energy scale. These equations 
can be used to describe non-perturbative phenomena. A widely used form of the ERG is the Wetterich equation \cite{WETTERICH199390}, which describes the scale dependence of the effective average action:

\begin{equation}
\label{eq:wetterich}
    \dot{\Gamma}_k = \frac{1}{2}\text{STr}\biggl[\dot{R}_k(\Gamma^{(2)}_k+R_k)^{-1}  \biggr]
\end{equation}
where the dot is an abbreviation for the operation $k\partial_k$. The functional $\Gamma_k$ is the Legendre transform of the generating functional of the connected Green functions plus a scale-dependent mass term, called the regulator function $R_k$, and $\Gamma^{(2)}_k$ is the inverse propagator containing the physical mass. All the different formulations of the ERG equations require some sort of regularization. The regulator vanishes in the low energy limit of the theory. The super-trace contracts all momenta and group indices, therefore this equation can be viewed as a one-loop expression with an operator insertion ($\dot{R}_k$) and no external legs. 
The functional $\Gamma_k$ possesses the linear symmetries of the original Lagrangian if the regulator also does. 
In order to solve Eq.~(\ref{eq:wetterich}), one has to make an ansatz for $\Gamma_k$ comprised of a finite number of functions, consistent with the symmetries of the original theory, and specify the regulator function.

A widely used approach in terms of the ansatz is the derivative expansion. In this method, the leading-order (or local potential approximation, LPA) only has a scale-dependent potential and a canonical kinetic term. An important feature of the exact renormalization group is that even the irrelevant couplings acquire non-trivial scale dependence during the RG-flow. This observation leads one to believe that the LPA prediction can be improved by including couplings, corresponding to scale dependent functions, which multiply all operators but the unit operator. Consequently, the next-to-leading order (NLO) introduces scale dependent functions multiplying every independent operator with two derivatives. Similarly, at the NNLO operators with four derivatives appear. 
This expansion makes the functional space of $\Gamma_k$ less and less truncated order by order and at the same time increases the number of terms in the truncated ansatz. 
One expects, that including higher orders in the derivative expansion improves the quality of the physical predictions. In fact, the convergence of this method has been demonstrated in Ref.~\cite{PhysRevLett.123.240604} up to N${}^3$LO
for the $Z_2$ universality class.

The dependence on the regulator is expected to vanish in the low-energy limit, $k\to 0$. As we study the critical theory, which is scale independent, we expect our physical predictions to be independent of the specific form of the regulator $R_k$. This is strictly true only if we do not truncate the functional space. The dependence of the physical predictions and the magnitude of this spurious dependence on the regulator is somewhat similar to the renormalization scale dependence in the perturbative quantum field theory.

\section{%Generic way of d
Deriving the beta-functions}
\label{sec:beta}

The system is critical in the Wilson-Fisher fixed point, which is the non-trivial solution of the fixed-point equation of the $\beta$-functions. We need to obtain the $\beta$-functions and the Wilson-Fisher fixed point to compute the critical exponents. The derivation of these $\beta$-functions is comprised of four steps for a given ansatz: (i) splitting the field to homogeneous and fluctuating pieces,
(ii) functional Taylor expansion of Eq.~(\ref{eq:wetterich}) in powers of the fluctuating field, (iii) expansion in the momenta corresponding to the fluctuating field, and finally (iv) classification and sorting of the different types of loop integrals, called threshold integrals. We automated these steps in a \texttt{Mathematica} code attached in a supplement.

\subsection{Functional and momentum expansions}

As an example, let us consider the ansatz for the $Z_2$ symmetric scalar model at
the NLO of the DE:
\begin{equation}
    \Gamma_k[\phi] = \frac{1}{2}\int_x Z_k(\rho_x)(\partial \phi_x)^2 +\int_x U_k(\rho_x),
\end{equation}
    where $\rho_x=\phi_x^2/2\equiv\phi(x)^2/2$, $\int_x \equiv \int\!\rd^d x$ (and simialrly $\int_p = (2 \pi)^{-d}\int\!\rd^d p$, to be used later), and $(\partial f)^2 \equiv (\partial_\mu f) (\partial^\mu f)$ for any $f$. The flow for $U_k$ is obtained by setting the field $\phi$ to be homogeneous $\phi_x= \Phi$ (meaning $\partial \Phi=0$) and solving Eq~(\ref{eq:wetterich}). In order to find $\dot{Z}_k(\rho\equiv \Phi^2/2)$ however, we expand Eq.~(\ref{eq:wetterich}) in terms a fluctuating field $\eta_x$ around a constant background $\phi_x = \Phi+ \eta_x$ and collect the terms proportional to $\mathcal{O}(\eta^2)$. In momentum space, this is given by
\begin{equation}\label{eq:nlobeta}
\bsp
    \int_Q &\biggl(\dot{Z}_k(\rho)Q^2 + \dot{U}_k'(\rho) + 2 \rho \dot{U}_k''(\rho) \biggr) \eta_Q \eta_{-Q} =
\\&= 
      \int_{p,r} \dot{R}_k(p^2) \mathcal{G}(p^2)(\eta \Gamma^{(3)})_{p,-r}\mathcal{G}(r^2)(\eta \Gamma^{(3)})_{r,-p}\mathcal{G}(p^2)
\\&
     -\frac{1}{2}\int_p \dot{R}_k(p^2) \mathcal{G}(p^2)(\eta \Gamma^{(4)} \eta)_{p,-p}\mathcal{G}(p^2),
\esp
\end{equation}
with $\mathcal{G}(p^2)$ being the regularized propagator ($(\Gamma^{(2)}_k+R_k)^{-1}$), $r=p\pm Q$
, and 
\begin{equation}
\bsp
&
(\eta \Gamma^{(3)})_{p,q} =   \eta_{-p-q} \frac{\delta^{(3)}\Gamma}{\delta \phi_p \delta \phi_q \delta \phi_{-p-q}}\biggr|_{\phi_x=\Phi},
\\&
(\eta \Gamma^{(4)} \eta)_{p,q} =  \int_{Q} \eta_{Q} \frac{\delta^{(4)}\Gamma}{\delta \phi_p \delta \phi_q \delta \phi_{Q} \delta \phi_{-Q} }\biggr|_{\phi_x=\Phi}\eta_{-Q}.
\esp
\end{equation}
Generally, in order to find $\dot{F}$, where $F$ multiplies an operator with $n$ derivatives one has to collect terms proportional to $\mathcal{O}(\eta^n)$. We denote the momentum of the fluctuating field $\eta$ with $Q$ for transparency. In case, there are multiple $\eta$ fields in the same expression their momenta are denoted with $Q_1$, $Q_2$ and so on.

The left hand side of Eq.~(\ref{eq:nlobeta}) shows, that in order to obtain $\dot{Z}_k(\rho)$, we have to expand the right hand side in $Q_\mu$ up to $Q^2$ and finally, identify the terms proportional to $Q^2$ as the beta function of $Z_k(\rho)$. The computations become naturally more complicated at NNLO, since then there are multiple momenta $Q_i$. For the sake of concreteness, the complete ansatz for the $Z_2$ symmetric scalar model at the fourth order of the derivative expansion reads as
\begin{equation}\label{eq:z2-nnlo}
\bsp
&
\Gamma_k[\phi] = \frac{1}{2}\int_x Z_k(\rho_x)(\partial \phi_x)^2 +\int_x U_k(\rho)
\\&+ 
\frac{1}{2}\int_x W_k(\rho_x)(\partial_\mu\partial_\nu \phi_x)^2
\\&+ 
\frac{1}{2}\int_x H_k(\rho_x)\phi_x (\partial \phi_x)^2(\partial^2 \phi_x)  
\\&+ 
\frac{1}{2}\int_xJ_k(\rho_x) (\partial \phi_x)^4
\esp
\end{equation}
This form has been studied in great detail without and also with expansion in the fields \cite{PhysRevLett.123.240604,dela2003}. The scale dependent functions $W_k$, $H_k$ and $J_k$ are obtained from $\Gamma_k$ via 
\begin{eqnarray}\label{eq:momdif}
W_k(\rho) &=& \lim_{Q_1 \to 0} \biggl(\frac{\partial}{\partial Q_1^2}\biggr)^2\Gamma^{(2)}_{Q_1,Q_2},
\\
H_k(\rho) &=& -\frac{1}{2\Phi}\lim_{Q_1,Q_2 \to 0} \frac{\partial}{\partial Q_1^2}\frac{\partial}{\partial Q_2^2}\Gamma^{(3)}_{Q_1,Q_2,Q_3},
\\
J_k(\rho) &=& -\frac{1}{4}\lim_{Q_1,Q_2,Q_3 \to 0} \frac{\partial}{\partial Q_1^2}\frac{\partial}{\partial (Q_2\cdot Q_3)}\Gamma^{(4)}_{Q_1,Q_2,Q_3,Q_4},
\end{eqnarray}
as the coefficients of the integrands in the integrals $\int_{Q_1,\dots,Q_n}\! \prod_{i=1}^n \eta_{Q_i} \delta(\sum_{i=1}^n Q_i)$ for $n=2$, 3 and 4.
Note, that the scale dependent functions can be acquired by any permutation of the momentum indices $Q_i$ in the differentiation. 

The $O(N)$ symmetric models introduce an additional index on the field corresponding to the symmetry group and can be generalized from the $Z_2$ symmetric models in a straightforward way.
The complete ansatz used in this work is given in Eq.~(\ref{eq:fullon}). A slightly different, but equivalent ansatz is used in Ref.~\cite{full-on}.
\subsection{
Threshold integrals
}\label{sec:threshint}

After sorting the different types of $\int_p$ integrals that appear in the  formula of a general $\dot{F}_k$ in the $Z_2$ symmetric model at NNLO, one finds three such types:
\begin{eqnarray}
\label{eq:master-integrals}
 L_m^{d+a} &=& \int_p p^a\frac{\dot{R}_k(p^2)}{G(p^2)^m}, \\
 M_{m,b}^{d+a,\beta} &=& \int_p p^a(\partial_{p^2}^{\beta} G(p^2))^b\frac{\dot{R}_k(p^2)}{G(p^2)^m},\\
 N_{m,b,c}^{d+a,\beta,\gamma} &=& \int_p p^a(\partial_{p^2}^{\beta}
 G(p^2))^b(\partial_{p^2}^{\gamma} G(p^2))^c\frac{\dot{R}_k(p^2)}{G(p^2)^m}
\end{eqnarray}
where $m$, $b$, $c$, $\beta$ and $\gamma$ are positive integers and $a$ is a non-negative one. 
We have also introduced $G(p^2)$ as the regularized inverse propagator ($(\Gamma^{(2)}_k+R_k)$). As we consider the NNLO of the DE, derivatives of the inverse propagator appear up to the fourth derivative. This yields the constraint $b\beta+c\gamma \leq 4$ for the threshold integral parameters.

In the  $O(N)$ symmetric models two types of propagators appear: one massive and one corresponding to the $N-1$ Goldstone modes. This proliferates the types of threshold integrals.

\subsection{Regulator functions}\label{sec:regulator}
 The regulator itself is a function of the loop momentum squared $p^2$ and the running scale $k$. It is usually expressed as the function of the dimensionless ratio $y=p^2/k^2$:
\begin{equation}
    R_k(p^2) = Z_k k^2 y~r(y), 
\end{equation}
where the explicit form of the regulator is defined by the function $r(y)$, $Z_k=1$ at LPA and $Z_k\equiv Z_k(\rho=\rho^*)$ at higher orders of the DE with $\rho^*$ being a reference value, detailed in Sect.~\ref{sec:polyexp}.  In general the form of the regulator is very flexible, yet it has to obey some requirements \cite{WETTERICH199390}. 

In order to obtain numerical results, one has to specify the regulator function.
In this work we use two different types. The $\Theta_2$-regulator introduced in Ref.~\cite{Litim2001} reads as
\begin{equation}\label{eq:regulator}
    r_{\Theta}(y) = \alpha \frac{(1-y)^2}{y} \Theta(1-y)
\end{equation}
where $\Theta(x)$ is the Heaviside step function.
The regulator (\ref{eq:regulator}) is the simplest possible regulator which can be used in $\partial^4$-order calculations. The caveat is that it is not applicable beyond $\partial^4$-order due to the appearance of undefined Dirac-delta functionals $(\delta(0))$ in the final equations.
Generally, at $\partial^n$-order the integral containing the highest $G$-derivative is:
\begin{equation}
M_{m,1}^{d,n} = \frac{\Omega_d}{(2\pi)^d} k^d\int\!\rd y\, y^{-1+d/2}(\partial_y^{n}G(y)) \frac{\dot{R}_k(y)}{G(y)^m},
\end{equation}
where we have changed to the variable $y=p^2/k^2$. For the regulator (\ref{eq:regulator}) and $n=4$ this integral takes the form
\begin{equation}\label{eq:ambiguous}
\bsp
 M_{m,1}^{d,4} &= -4\alpha^2 \frac{\Omega_d}{(2\pi)^d}(Z_k^2 k^{d+2}) \times
\\&\times
 \int\!\rd y  \frac{y^{-1+d/2}(y^2-1)\Theta(1-y)}{G(y)^m}\delta'(1-y)
\\& \equiv
4 \alpha^2 (Z_k^2 k^{d+2}) \frac{\Omega_d}{(2\pi)^d} \frac{1}{G(1)^m}.
\esp
\end{equation}
This integral is ambiguous in the sense, that the result is obtained by integration by parts and then defining $\Theta(0)$ to be $1/2$. This ambiguity is lifted, when one considers (\ref{eq:regulator}) as the limit of a $C^\infty$-type regulator function, such as (\ref{eq:betareg}). The process to do so is detailed in App.~\ref{app:theta-reg}.
The integrals, which contain $\partial_y^{3}G_k(y)=-2\alpha (Z_k k^2)\delta(1-y)$ vanish, because the distributional product $x\delta(x)$ is zero and every integral contains $(1-y)$ through $\dot{R}_k(y)$.

The second regulator we use here is called the exponential regulator
\begin{equation}\label{eq:exp-regulator}
    r_{exp}(y) = \alpha \frac{e^{-y}}{y},
\end{equation}
which is a $C^\infty$ function and has the advantage over the regulator containing the $\Theta$-function that it can be used at any orders of the derivative expansion. Both $r_\Theta$ and $r_{exp}$ remain unchanged in the $Z_2$ and $O(N)$ symmetric scalar models.

We vary the value of $\alpha$ and compute its effect on the critical exponents. We consider the extrema of these functions as the optimal values in our final predictions. This is the implementation of the the principle of minimal sensitivity (PMS) \cite{PMS-opt,dela-nlo}. In practice, we locate the Wilson-Fisher fixed point for a fixed regulator for several values of $\alpha$, which simultaneously yields $\eta(\alpha)$ as the anomalous dimension is just a function of the couplings in the model. In each case we applied the PMS $\eta(\alpha)$ is either an upside or downside facing paraboloid. The optimal value of $\eta^{opt}$ is the minimum/maximum of this paraboloid at $\alpha^{opt}$ and we accept $\nu(\alpha^{opt})$ and $\omega(\alpha^{opt})$ as $\nu^{opt}$ and $\omega^{opt}$. In this sense, we only apply the PMS on the anomalous dimension.

\subsection{Polynomial expansion and exponents}\label{sec:polyexp}
In order to compute the critical exponents one has to use dimensionless quantities. The mass dimension of some are given as
\begin{equation}
     [\phi] = (d-2+\eta_k)/2,~~~ [U]=d,
\end{equation}
where $\eta_k$ is the running anomalous dimension, which is defined by 
\begin{equation}
    \eta_k = - k~\partial_k \ln Z_k(\rho^*).
\end{equation}
The running anomalous dimension becomes the critical exponent $\eta$ in the fixed point. The Euclidean dimension $d$ is a continuous parameter in the beta functions of the dimensionless couplings. We set its value to $d=3$ throughout this work. The beta functions for the dimensionless scale dependent functions are partial differential equations with the scale $k$ and the dimensionless field $\tilde{\rho}$ (we denote the dimensionless quantities with tilde) as independent variables. One strategy to solve these equations is to Taylor expand the dimensionless scale dependent functions in power of the dimensionless field around a reference point $\rho^*$
\begin{equation}
\label{eq:beta-functional}
   \tilde{F}_k(\tilde{\rho)} = \sum_{n=0}^{M_F}  \frac{\tilde{f}_n(k)}{n!}(\tilde{\rho}-\rho^*)^n.
\end{equation}
This reduces the coupled set of partial differential equations to a coupled set of ordinary differential equations. 
This course of action has been taken for example in Refs.~\cite{dela-nlo,dela2003}.
There are two well known choices for $\rho^*$. It can either be zero $(\rho^*=0)$ or the running minimum $\rho^* = \kappa_k$ of the most basic scale dependent function, the local potential $U_k$. Throughout this work we use $\rho^* = \kappa_k$, because it provides a faster convergence of the physical results with increasing $M_F$ than expanding around the vanishing field \cite{poly-stability,LITIM2002128}. 
We denote the highest power in the Taylor series of a general scale dependent function $F_k$ with $M_F$, if the subscript contains multiple capital Latin letters such as $M_{WHJ}$, it means that the scale dependent functions $W_k$, $H_k$ and $J_k$ are truncated at identical powers $M_W=M_H=M_J\equiv M_{WHJ}$.

The Wilson-Fisher fixed point is the nontrivial fixed point solution of the beta functions. Once it is located, the critical value of the anomalous dimension $\eta$ is determined. The critical exponent of the correlation length $\nu$ and its subleading scaling corrections $\omega,\omega_i$ are obtained by linearizing the RG-flow in the vicinity of the fixed point. The eigenvalues of the Jacobian matrix $J_{ij}=\partial \beta_{\tilde{g}_i}/\partial \tilde{g}_j$, with $\tilde{g}_i$ being a general dimensionless coupling from the model, at the fixed point are $-\nu^{-1}<\omega<\omega_1<\ldots$ in increasing order. 

The polynomial expansion gives very good predictions at $d=3$ as demonstrated in Ref.~\cite{dela2003}. However, this might not be the case for $d<3$. As $d$ is lowered, new couplings $g_n$ corresponding to the vertex $\phi^{2n}$ become marginal ($[g_n]=0$) at $n=d/(d-2)$. If $g_{n+1}$ is marginal, then $g_n$ is relevant. At $d=4$ only the mass squared is a relevant coupling  ($[g_1]>0$) and the quartic interaction is marginal ($[g_2]=0$). At $d=3$ there are two relevant couplings ($[g_1]>0$ and $[g_2]>0$) and thus a nontrivial fixed point, the Wilson-Fisher fixed point appears. At $d=8/3$ the coupling $g_3$ also becomes relevant and introduces a new nontrivial fixed point besides the Wilson-Fisher one. This makes finding the Wilson-Fisher fixed point much more difficult. In particular, in Ref.~\cite{polyexp-bound} it has been found that the Euclidean action is not bounded from below in the fixed point, which sets a bound on the applicability of the polynomial expansion.
\section{Wynn's epsilon algorithm}\label{sec:wynn}
In many instances, the prediction of an exponent $X$ at successive orders of the DE, $X_{LPA}$, $X_{NLO}$, $X_{NNLO}$ and so on, form a convergent series alternating around the exact value $X$. This has been discussed in great detail in Ref.~\cite{PhysRevLett.123.240604}. In Ref.~\cite{full-on} the authors use the small parameter $1/4~-~1/9$ of the DE to improve their predictions on the critical exponents of the $O(N)$ symmetric scalar models at NNLO of the DE.

One may also turn to a similar, yet different approach to improve exponent predictions in the derivative expansion. Several series acceleration methods exist and are used successfully to accurately compute the limit of a slowly converging sequence. One of the most robust of these algorithms is Wynn's epsilon algorithm \cite{wynn1,wynn2}. It is already applicable if one only has the first three elements $a_1,a_2,a_3$ of a sequence $(a_n)$. In that case, the third element is improved as
\begin{equation}
\tilde{a}_3=a_2+\frac{1}{-\frac{1}{-a_1+a_2}+\frac{1}{-a_2+a_3}}= \frac{-a_2^2+a_1 a_3}{a_1-2a_2+a_3}.   
\end{equation}
Given the critical exponent $X$, this means, that the improved prediction of the DE is
\begin{equation}
\tilde{X}=\frac{-X_{NLO}^2+X_{LPA} X_{NNLO}}{X_{LPA}-2X_{NLO}+ X_{NNLO}}.   
\end{equation}
The formula is even simpler for the anomalous dimension as the LPA prediction for it is zero.
We employ Wynn's algorithm when it works the best, i.e.~with alternating sequences. The ERG predictions for the $O(N)$ critical exponents at different orders of the DE show that while the predictions for $\nu$ and $\eta$ do show an alternating behavior, this is not always the case for $\omega$. Among the exponents we have computed this is the case for $\omega$ corresponding to the $O(2),O(3)$ and $O(4)$ symmetric models. In those instances we did not apply Wynn's $\epsilon$ algorithm, and only cited our NNLO predictions as our final value for $\omega$ for the $O(2),O(3)$ and $O(4)$ symmetric models.

We use this method to accurately extrapolate to higher orders of the DE
and thus obtain more precise predictions, since
the functional space of $\Gamma_k$ is less truncated at higher orders of the DE.
An other systematic source of error is that of the DE itself. If one insists on using Wynn's epsilon algorithm, then it is necessary to compute the $N^{3}LO$ prediction of the DE in order to give a conservative estimate on this error. In order to still give reliable predictions, we use the well-grounded error estimate for the DE proposed in Ref.~\cite{full-on} detailed in App.~\ref{app:error-estimate}.
\section{Predictions for the $Z_2$ symmetric scalar model}\label{sec:z2}
We derived the beta functions for the dimensionless scale dependent functions ($U_k,Z_k,W_k,H_k,J_k$) in the ansatz (\ref{eq:z2-nnlo}) using a \texttt{Mathematica} code. We verified the correctness of $\dot{U}_k$ and $\dot{Z}_k$ (at $\partial^2$-order) to be the same as in the literature \cite{TETRADIS1994541,dela-nlo}. We expanded these functions in the powers of the field yielding the beta functions for the dimensionless couplings $\tilde{f}_n(k)$ in Eq.(\ref{eq:beta-functional}). We have calculated the effect of increasing $M_F$ on the exponents. We start with the LPA, where the only scale dependent function is $U_k$ and locate the Wilson-Fisher fixed point with truncation threshold $M_U=4$. In the next step, we locate the fixed point for $M_U=5$ using the previous fixed point solution with $\tilde{u}_5=1$ as initial value. After this, we move on to $M_U=6$ using the previous fixed point solution with $\tilde{u}_6=1$ as initial value. In this iterative manner, we find the Wilson-Fisher fixed point for up to $M_U=8$. At the NLO, we have an additional scale dependent function $Z_k$ and nonzero anomalous dimension. We start with locating the fixed point at $M_U=8$ and $M_Z=0$, but including the effect of anomalous dimension and simply use the LPA values for $M_U=8$ as initial value. Next, we apply to $M_Z$ the iterative procedure used to find the fixed point for $M_U=8$ at the LPA. We find the Wilson-Fisher fixed point for up to $M_U=8$ and $M_Z=8$. At NNLO, we have three scale dependent functions $W_k$, $H_k$ and $J_k$. We start looking for the Wilson-Fisher fixed point at $M_U=M_Z=8$ with $M_W=M_H=M_J=0$, and setting the initial values to be $\tilde{w}_0=\tilde{h}_0=\tilde{j}_0=1$ for the new couplings. Finally, we also apply here the previously described iterative algorithm but we increase simultaneously  $M_W$, $M_H$ and $M_J$ and denote this value with $M_{WHJ}$. The upper limit where we have located the Wilson-Fisher fixed point is $M_{WHJ}=7$.

We have computed the fixed points with the two regulators discussed in Sect, \ref{sec:beta}. Using (\ref{eq:regulator}) with $\alpha = 1/2$ reduces the integrals (\ref{eq:master-integrals}) to linear combinations of the ${}_2F_1$ hypergeometric function, which greatly increases the speed of computations compared to (\ref{eq:exp-regulator}) with any value of $\alpha$.
\begin{figure}[ht]
\includegraphics[width=0.9\linewidth,height=100mm]{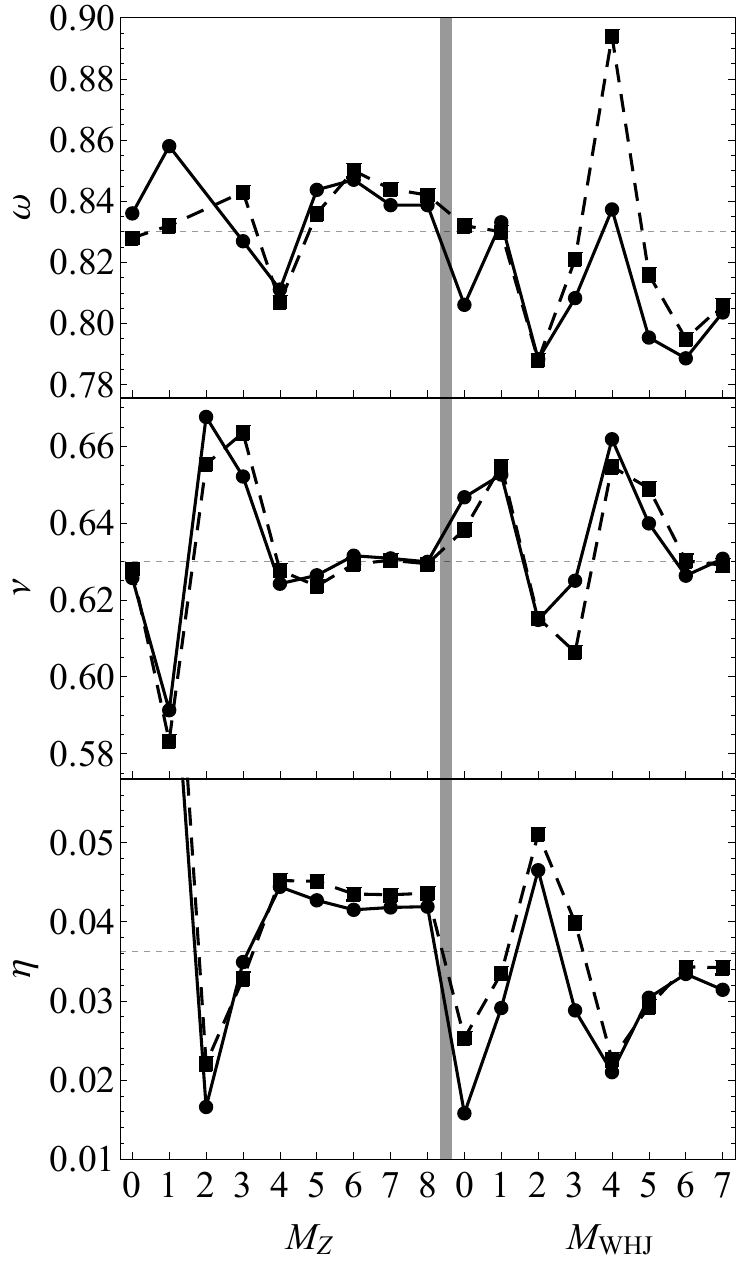}
\caption{\label{fig:full-o1-nu} The effect of the polynomial truncation in the $Z_2$ symmetric scalar model at NLO (left) and NNLO (right) on the critical exponents $\nu$, $\eta$ and $\omega$ at $M_U=8$. The continuous line with disks corresponds to the regulator $r_\Theta$ with $\alpha=1/2$, the dashed line with squares to the regulator $r_{exp}$ with $\alpha=1$. The CB values are shown for reference with the dotted horizontal line.}
\end{figure}

\begin{figure}[h]
\includegraphics[width=0.9\linewidth,height=100mm]{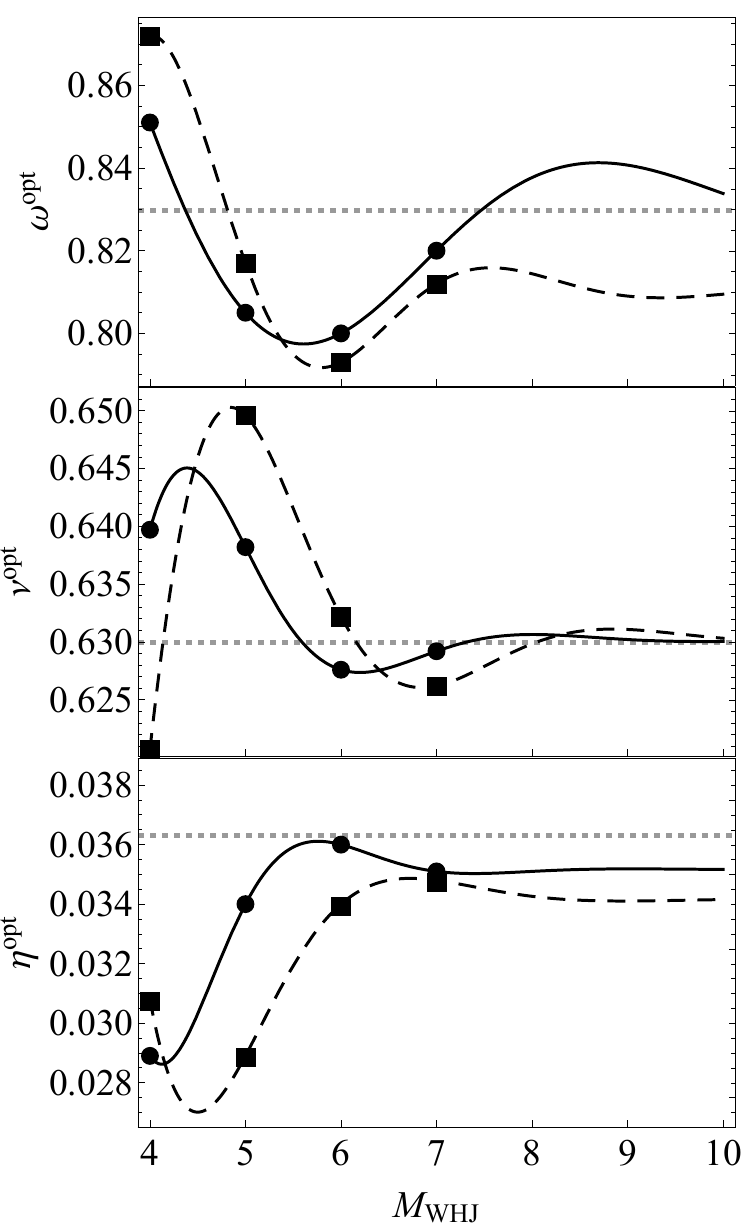}
\caption{\label{fig:full-o1-opt} A decaying function fit on the PMS optimized values of the exponents of the $Z_2$ symmetric scalar model at truncation $M_{WHJ} =4$ and above. The disks correspond to the values obtained with $r_\Theta$, the squares to the values obtained with $r_{exp}$. The dashed horizontal line shows the CB values.}
\end{figure}

The effect of the gradual inclusion of the new couplings can be seen on the left column of Fig.~\ref{fig:full-o1-nu}., which agrees with \cite{dela2003}. The most important conclusion is that while at $\partial^2$-order the contributions of the Taylor expansion in field variable become small for $M_Z>4$ this threshold power value at $\partial^4$-order is somewhat larger, $M_{WHJ}=6$. The magnitude of these contributions start to decrease monotonically for $M_Z>3$ at NLO and $M_{WHJ} > 4$ at NNLO. Next, we apply the principle of minimal sensitivity to $M_{WHJ}\geq 4$, which corresponds to the last four data points in each row of Fig. \ref{fig:full-o1-nu}. We have found that the optimal values $\alpha^{opt}$ for the regulators (\ref{eq:regulator}) and (\ref{eq:exp-regulator}) exhibit only small fluctuations around $\alpha^{opt}=0.35$ and $0.8$ for  $M_{WHJ}=0,..,7$. The only 
instance we have not found a PMS solution is for the truncation $M_{WHJ}=0$. The explicit values for the optimal parameter value $\alpha^{opt}$ corresponding to  $M_{WHJ}=7$ are found to be $\alpha^{opt}=0.30$ for the regulator (\ref{eq:regulator}) and  $\alpha^{opt}=0.76$ for (\ref{eq:exp-regulator}).
Once we acquire the optimized results in this asymptotic regime, where each successive contribution from the Taylor expansion is smaller than the previous one, we fit a decaying and alternating function to these data points in an attempt to capture the behavior of the Taylor series and resum the corrections from the Taylor expansion. The model function in every instance is
\begin{equation}\label{eq:resumtaylor}
X(M) = a+ b~\re^{-c~M} \sin(d~M + e),    
\end{equation}
with the independent variable being $M$ the degree of polynomial truncation and the fitted parameters are $a,b,d,e$ and $c>0$.
This step is shown in Fig.~\ref{fig:full-o1-opt}. We consider our findings to be the $M_{WHJ} \to \infty$ limit of these fitted functions, that is we identify the exponent as  $X(M\to\infty) =a$ from the model function (\ref{eq:resumtaylor}). We do not apply Wynn's epsilon algorithm here, because the corrections from increasing $M_{WHJ}$ is not a simple alternating series. In the asymptotic regime however, shown with the PMS optimized exponent on Fig.~\ref{fig:full-o1-opt}, these corrections alternate around their limiting value with periodicity of at least two. For instance, we expect that the correction from $M_{WHJ}=8$ increase the value of $\nu^{opt}$ compared to $M_{WHJ}=7$ and the higher corrections to have smaller effect than this. The model function (\ref{eq:resumtaylor}) takes this into account correctly.

Every beta function contains terms proportional to $\eta$ through $\dot{R}_k$. Considering only the exponents $\nu$ and $\eta$, the inclusion of these terms in $\dot{\tilde{U}}_k$ gives a $~1\%$ and $~5\%$ correction, while in $\dot{\tilde{Z}}_k$ they give $~0.1\%$ and $~0.5\%$ correction compared to not including those. We have also inspected the inclusion of these terms into $\dot{\tilde{W}}_k,\dot{\tilde{H}}_k$ and $\dot{\tilde{J}}_k$ for the truncation $M_U=8$ and $M_Z=8$ with $M_{WHJ}\leq 4$ and found that this characteristically gives a $~0.02\%$ and $~0.008\%$ correction to the exponents. We have neglected this correction in $\dot{\tilde{W}}_k,\dot{\tilde{H}}_k$ and $\dot{\tilde{J}}_k$ for $M_{WHJ}\geq 5$ and considered it as one source of uncertainty. The other source comes from the truncation of $\tilde{U}_k$ and $\tilde{Z}_k$. As a double check, we have computed the fixed point for truncation $M_U=9$, $M_Z=8$ and $M_U=9$, $M_Z=9$ at NLO. We have found that the inclusion of the coupling $\tilde{u}_9$ has negligible effect compared to the inclusion of $\tilde{z}_9$. Our final predictions for the critical exponents of the $Z_2$ symmetric model are shown in Tab.~\ref{tab:o1-summary}. 
The method to obtain the predictions and their corresponding uncertainty are detailed in App.~\ref{app:error-estimate}.
\begin{table}[h!]
    \centering
\begin{tabular}{ |p{2.3cm}||p{1.8cm} p{2cm} p{2cm}|  }
 \hline
 Method & $\nu$ & $\eta$ & $\omega$ \\
 \hline
 \hline
LPA & $0.6504(7)$    &  $0$ & $0.654(1)$   \\
NLO & $0.629(5)$ &  $0.042(11)$ & $0.84(4)$   \\
NNLO & $0.6302(4)$ &  $0.0347(30)$ & $0.820(10)$   \\
improved & $0.6301(4)$ &  $0.0358(30)$ & $0.822(10)$   \\
 \hline
 \hline
 $\partial^4$, field exp. & $0.632$ & $0.033$ & \\
\hline
 $\partial^6$, no field exp. & $0.63012(16)$ & $0.0362(12)$ & $0.832(14)$\\
\hline
 MC & $0.63002(10)$ & $0.03627(10)$ & $0.832(6)$\\
\hline
 six-loop PT & $0.6304(13)$ & $0.0335(25)$ & $0.799(11)$\\
\hline
$\epsilon^6$, epsilon exp. & $0.6292(5)$ & $0.0362(6)$ & $0.820(7)$\\
\hline
CB & $0.629971(4)$ & $0.0362978(20)$ & $0.82968(23)$\\
\hline
\end{tabular}
    \caption{Our findings for the exponents of the $Z_2$ symmetric scalar model in $d=3$ Euclidean dimensions (top four rows) for different orders of the DE and the improved, final prediction. The uncertainties are the sum of the uncertainties from the polynomial expansion and the regulator dependence. We compared these to some other methods: DE at NNLO $(\partial^4)$ with field expansion \cite{dela2003}, at N${}^3$LO $(\partial^6)$ without field expansion \cite{full-on}, MC \cite{MC-ising}, six-loop perturbation theory at fixed $d=3$ \cite{Guida_1998}, $d=4-\epsilon$ expansion at $\epsilon^6$ \cite{eps-expand} and the CB method \cite{bootstrap-ising}.
    \label{tab:o1-summary}
    }
\end{table}

\section{NNLO for the O(N) symmetric scalar models}\label{sec:on}

\subsection{Modifications compared to the $Z_2$ symmetric case}
There are more scale dependent functions in the $O(N)$ symmetric scalar model beyond the LPA than in the $Z_2$ symmetric one, due to an additional group index. At NLO, there are two instead of the one $Z_k$, but at NNLO the number of independent scale dependent functions increase to ten, compared to the three $W_k,H_k$ and $J_k$. The complete $\partial^4$-order ansatz is
\begin{equation}\label{eq:fullon}
\bsp
&
\Gamma_k[\vec{\phi}]=\int_x \biggl\{U_k 
+ \frac{1}{2}Z_k (\partial \phi^a_x)^2+\frac{1}{4}Y_k(\partial \rho_x)^2
\\&+ 
 \frac{1}{2} W_{1,k}(\partial_{\mu}\partial_{\nu}\phi^a_x)^2
+\frac{1}{4} W_{2,k}(\phi^a_x\partial_{\mu}\partial_{\nu}\phi^a_x)^2
\\&+
 \frac{1}{2} H_{1,k}(\partial \phi^a_x)^2(\phi^b_x\partial^2\phi^b_x)
+ H_{2,k}(\partial_\mu \rho_x)(\partial^\mu \phi^b_x)(\partial^2\phi^b_x)
\\&+
 \frac{1}{4} H_{3,k}(\partial \rho_x)^2(\phi^a_x\partial^2\phi^a_x)
+\frac{1}{8} J_{5,k} (\partial \rho_x)^4 
\\&+
 \frac{1}{2} J_{1,k} (\partial \phi^a_x)^2(\partial \phi^b_x)^2
+\frac{1}{2} J_{2,k} (\partial_\mu \phi^a_x)(\partial_\nu \phi^a_x)(\partial^\mu \phi^b_x)(\partial^\nu \phi^b_x)
\\&+
 \frac{1}{4} J_{3,k} (\partial \rho_x)^2(\partial \phi^a_x)^2
+\frac{1}{4} J_{4,k} (\partial_\mu \rho_x)(\partial_\nu \rho_x)(\partial^\mu \phi^b_x)(\partial^\nu \phi^b_x)\biggr\}.
\esp
\end{equation}
where $\vec{\phi}$ is the $N$ component scalar field and $\rho_x = \phi^a_x \phi^a_x/2$ is the invariant under the $O(N)$ symmetry transformation. We have suppressed the field dependence of the scale dependent functions in (\ref{eq:fullon}) to be more transparent. Due to the appearance of the Goldstone modes in addition to the one massive mode in the $Z_2$ symmetric model, we have two anomalous dimensions corresponding to these modes: 
\begin{eqnarray}\label{eq:on-eta}
    \eta &=& - k~\partial_k \ln Z_k(\rho^*) \\
    \tilde{\eta} &=& - k~\partial_k \ln \biggl(Z_k(\rho^*) + \rho^* Y_k(\rho^*)\biggr)\equiv - k~\partial_k \ln \tilde{Z}_k(\rho^*)\nonumber\\
\end{eqnarray}
These anomalous dimensions are equal in the critical point. In our numerical check, we use this fact to ensure the correctness of our equations. Besides the field, the regulator function also receives $O(N)$ indices. We choose
\begin{eqnarray}\label{eq:on-regulator}
 R_{k}^{ab}(y) = \delta^{ab} Z_k(\rho^*) k^2 y~r(y)
\end{eqnarray}
where $\delta^{ab}$ is the Kronecker-delta matrix, such that the regulator mass matrix is already diagonalized in the $O(N)$ space. In order to facilitate the bookkeeping of the $O(N)$ indices, we introduce projectors $P_A^{ab}$ with ($A=\parallel,\perp$) to the radial ($P_\parallel^{ab}=e^a e^b$) and perpendicular (Goldstone) ($P_\perp^{ab}=\delta^{ab} - e^a e^b$) directions in the $O(N)$ space, with $e^a$ being the unit vector.
\begin{figure}[hbt!]
\includegraphics[width=0.9\linewidth,height=100mm]{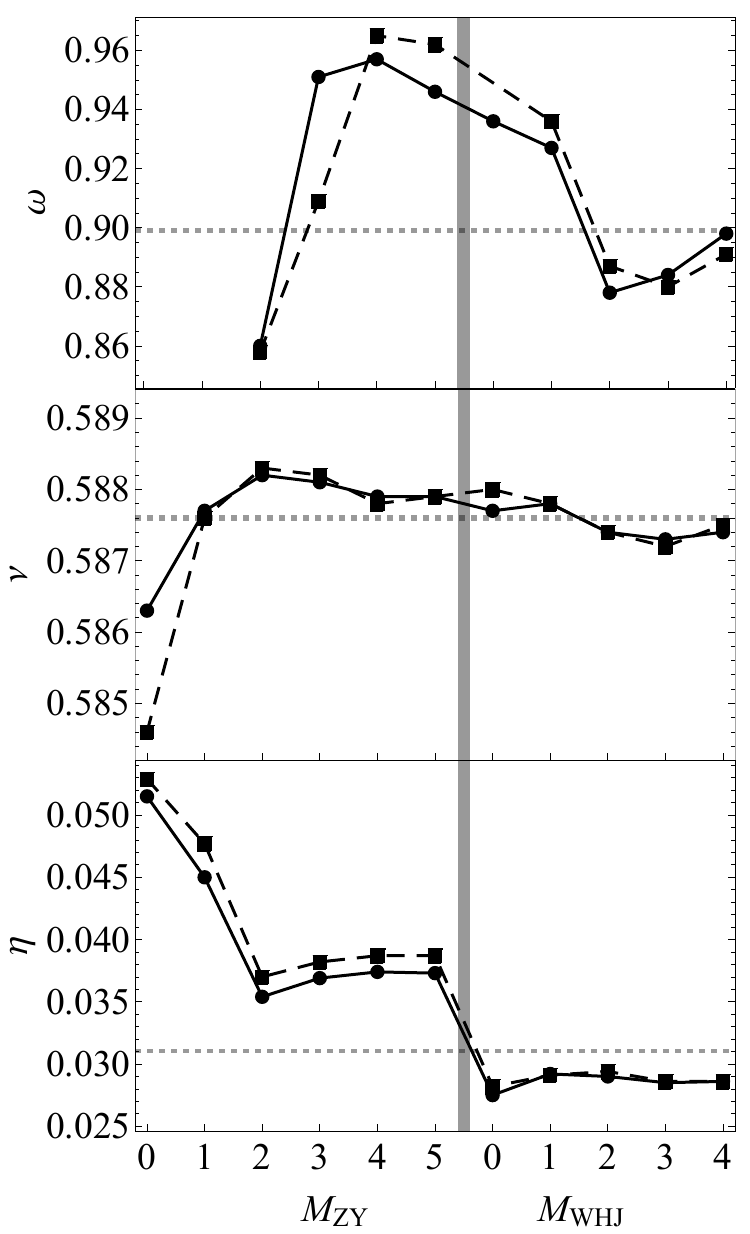}
\caption{\label{fig:full-o0-field}The dependence of the critical exponents $\nu$, $\eta$ and $\omega$ on the order of polynomial truncation for the $O(0)$ symmetric model at $M_U=8$. The vertical line separates our NLO results (left) from the NNLO ones (right). The dotted horizontal line shows the corresponding MC result. The continuous curve with disk markers belongs to the $\Theta$-type regulator (\ref{eq:regulator}) with $\alpha=1/2$, while the dashed curve with rectangle markers belong to the exponential-type regulator (\ref{eq:exp-regulator}) with $\alpha=1$. At the points, where $\omega$ is not shown, it is a complex number.}
\end{figure}
The scale dependent functions $Y_k$, $W_{i,k}$, $H_{i,k}$ and $J_{i,k}$ are obtained by the same momentum derivatives (Eq.~(\ref{eq:momdif})) as $Z_k$, $W_k$, $H_k$ and $J_k$ in the $Z_2$ symmetric model as coefficients of the integrands in $\int_{Q_1,\dots,Q_n}\!\prod_{i=1}^n \eta_{Q_i}^{A_i}\delta(\sum_{i=1}^n {Q_i})$. The capital Latin letters correspond to either $\parallel$ or $(\perp,a)$. Using the projectors defined above one has
\begin{eqnarray}
P_\parallel^{ab}\eta_x^a = \eta_x^\parallel ~~~\text{and}~~~P_\perp^{ab}\eta_x^a = \eta_x^{\perp,a}.
\end{eqnarray}
In this method, every $O(N)$ index is contracted in the final result, so that $\eta_{Q}^{\perp,a}$ may occur only in pairs, such as $\eta_{Q}^{\perp,a}\eta_{-Q}^{\perp,a}$. For instance, the left-hand side the Wetterich equation for $\mathcal{O}(\eta^2)$ Eq.~(\ref{eq:nlobeta}) modifies to
\begin{equation}
\bsp 
&
\int_Q \eta_{Q}^{\perp,a}\eta_{-Q}^{\perp,a} \biggl(\dot{Z}_k Q^2 + \dot{W}_{1,k} Q^4 + \dot{U}' \biggr)
\\&+
\int_Q \eta_{Q}^{\parallel}\eta_{-Q}^{\parallel} \biggl((\dot{Z}_k + \dot{Y}_k) Q^2 + (\dot{W}_{1,k} + \dot{W}_{2,k})Q^4
\\&~~~~~~~~~~~~~~~~~~~~
+\dot{U}' + 2 \rho \dot{U}''\biggr)
\esp
\end{equation}
with the ansatz in Eq.~(\ref{eq:fullon}).

We have followed the same steps of numerical analysis as we did for the $Z_2$ symmetric model. The system of $\beta$-functions are generated by a \texttt{Mathematica} code, which are then verified to reproduce the $\partial^2$-order results \cite{wette-on}. We applied the same iterative algorithm to find the Wilson-Fisher fixed point for high values of truncation $M$ as for the $Z_2$ symmetric model. At the LPA, we have computed the exponents for up to $M_U=8$. In the NLO we have increased simultaneously the truncation $M_Z$ of $Z_k$ and $M_Y$ of $Y_k$ for up to $M_Z=M_Y=5$ and denote this with $M_{ZY}$. At NNLO, we have ten scale dependent functions. In order to make it easier to find the Wilson-Fisher fixed point, we further divide the iterative algorithm to three parts. First, we locate the fixed point for the truncation $M_U=8$, $M_{ZY}=5$, $M_{W_1}=M_{W_2}=0$ with the initial values $\tilde{w}_{1,0}=\tilde{w}_{2,0}=1$. In the next step, we use this fixed point as initial value with $\tilde{h}_{1,0}=\tilde{h}_{2,0}=\tilde{h}_{3,0}=1$ for the truncation $M_U=8$, $M_{ZY}=5$, $M_{W_1}=M_{W_2}=0$ and $M_{H_1}=M_{H_2}=M_{H_3}=0$. In the last step we locate the fixed point with $M_{J_i}=0$ ($i=1,..,5$) also included. We denote this truncation with $M_{WHJ}=0$ when all the NNLO level scale dependent functions are included with zeroth order truncation in their Taylor expansion. We have computed the exponents for up to $M_U=8$, $M_{ZY}=5$ and $M_{WHJ}=4$.
\begin{figure}[hbt!]
\includegraphics[width=0.9\linewidth,height=100mm]{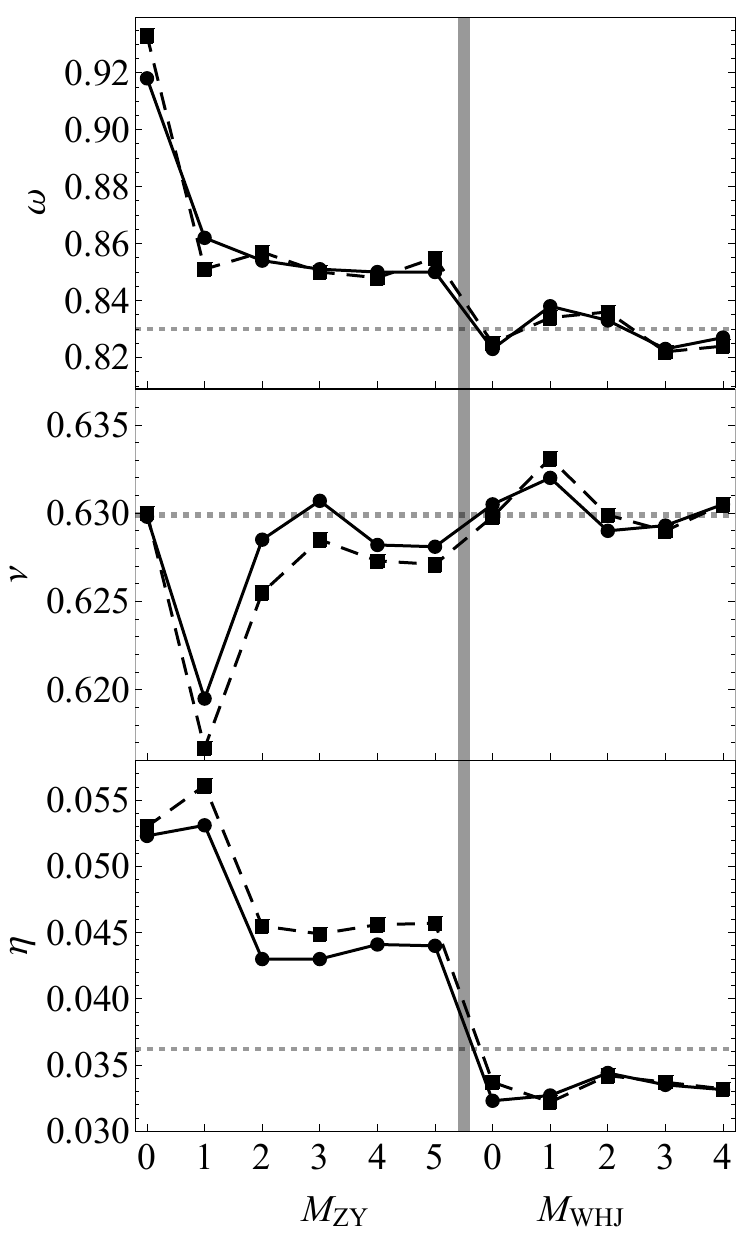}
\caption{\label{fig:full-o1-field}The dependence of the critical exponents $\nu$, $\eta$ and $\omega$ on the order of polynomial truncation for the $O(1)$ symmetric model at $M_U=8$. The vertical line separates our NLO results (left) from the NNLO ones (right). The dotted horizontal line shows the corresponding CB result. The continuous curve with disk markers belongs to the $\Theta$-type regulator (\ref{eq:regulator}) with $\alpha=1/2$, while the dashed curve with rectangle markers belong to the exponential-type regulator (\ref{eq:exp-regulator}) with $\alpha=1$.}
\end{figure}
\subsection{Numerical findings}\label{sec:on-num}

We have computed the critical exponents for the regulators (\ref{eq:regulator}) and (\ref{eq:exp-regulator}). The former one with $\alpha=1/2$ reduces a large number of the threshold integrals to ${}_2F_1$-type hypergeometric functions. This yields a significant speed boost in the computations compared to (\ref{eq:exp-regulator}) with any value of $\alpha$.

\begin{figure}[hbt!]
\includegraphics[width=0.9\linewidth,height=100mm]{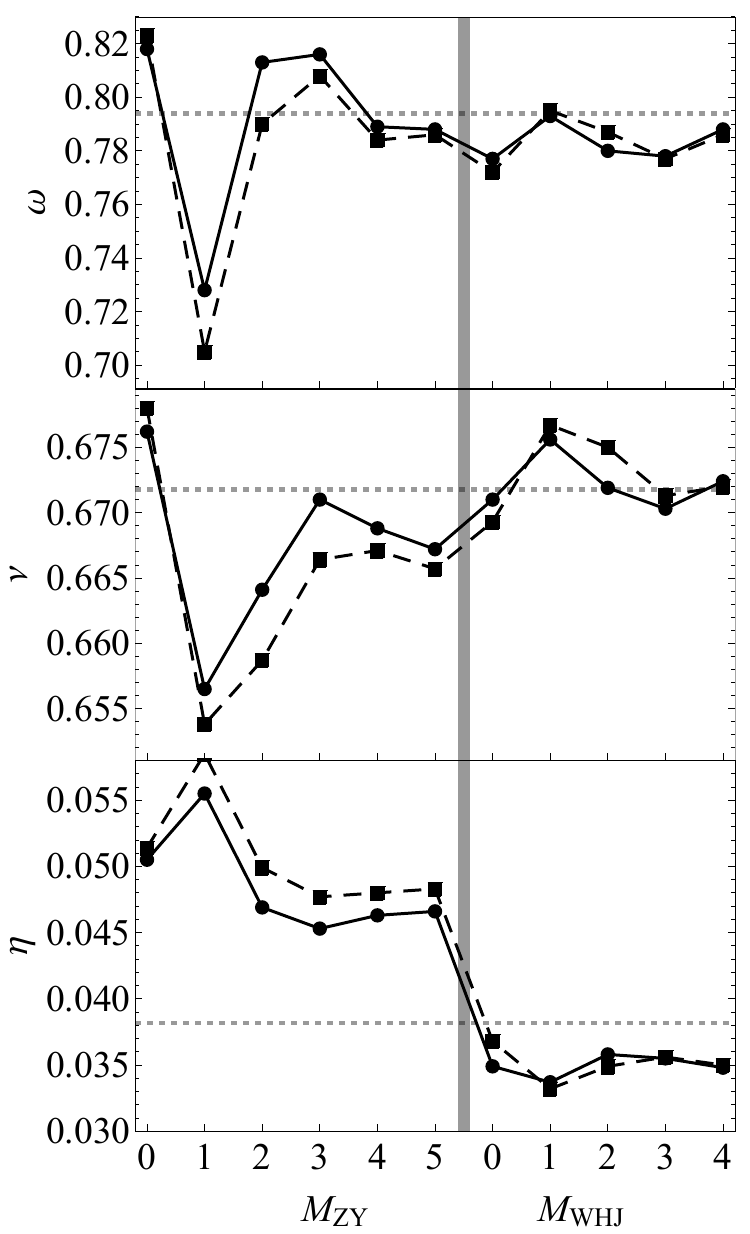}
\caption{\label{fig:full-o2-field}The dependence of the critical exponents $\nu$, $\eta$ and $\omega$ on the order of polynomial truncation for the $O(2)$ symmetric model at $M_U=8$. The vertical line separates our NLO results (left) from the NNLO ones (right). The dotted horizontal line shows the corresponding CB result. The continuous curve with disk markers belongs to the $\Theta$-type regulator (\ref{eq:regulator}) with $\alpha=1/2$, while the dashed curve with rectangle markers belong to the exponential-type regulator (\ref{eq:exp-regulator}) with $\alpha=1$.}
\end{figure}

\begin{figure}[hbt!]
\includegraphics[width=0.9\linewidth,height=100mm]{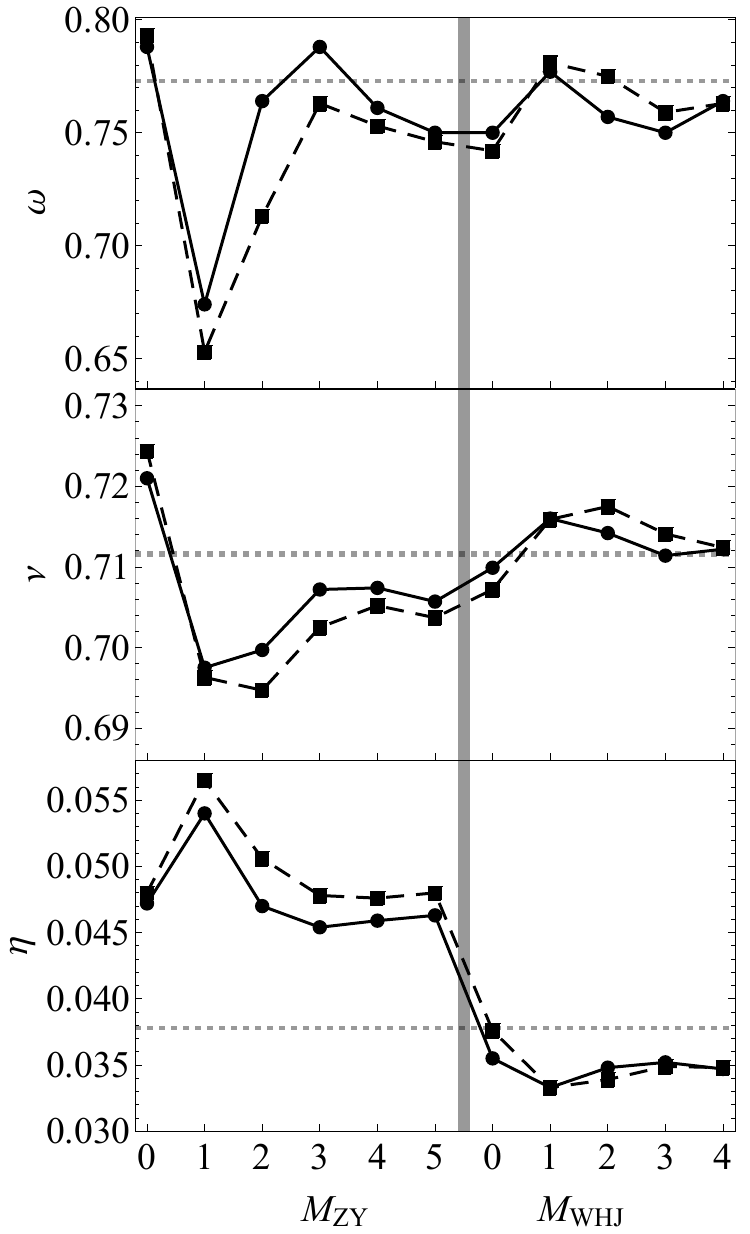}
\caption{\label{fig:full-o3-field}The dependence of the critical exponents $\nu$, $\eta$ and $\omega$ on the order of polynomial truncation for the $O(3)$ symmetric model at $M_U=8$. The vertical line separates our NLO results (left) from the NNLO ones (right). The dotted horizontal line shows the corresponding MC result. The continuous curve with disk markers belongs to the $\Theta$-type regulator (\ref{eq:regulator}) with $\alpha=1/2$, while the dashed curve with rectangle markers belong to the exponential-type regulator (\ref{eq:exp-regulator}) with $\alpha=1$.}
\end{figure}

\begin{figure}[hbt!]
\includegraphics[width=0.9\linewidth,height=100mm]{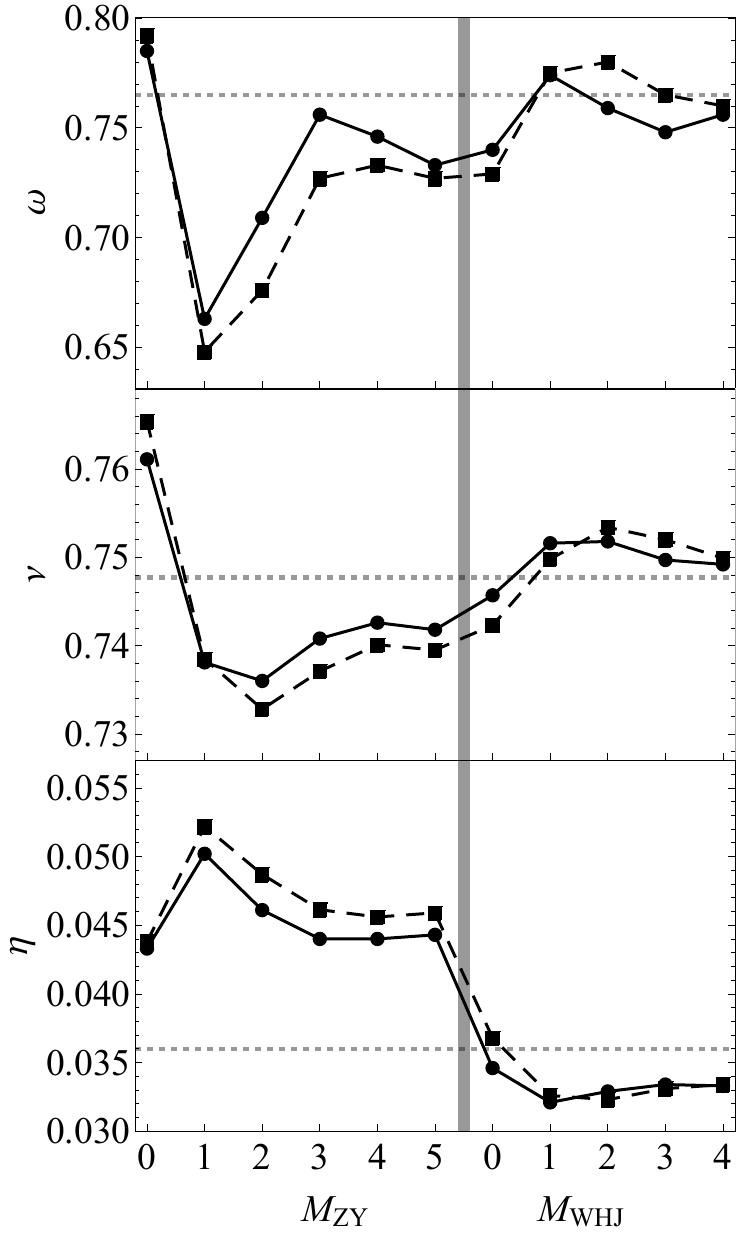}
\caption{\label{fig:full-o4-field}The dependence of the critical exponents $\nu$, $\eta$ and $\omega$ on the order of polynomial truncation for the $O(4)$ symmetric model at $M_U=8$. The vertical line separates our NLO results (left) from the NNLO ones (right). The dotted horizontal line shows the MC bootstrap result. The continuous curve with disk markers belongs to the $\Theta$-type regulator (\ref{eq:regulator}) with $\alpha=1/2$, while the dashed curve with rectangle markers belong to the exponential-type regulator (\ref{eq:exp-regulator}) with $\alpha=1$.}
\end{figure}

The effect of the gradual inclusion of the new couplings for the $O(N)$ symmetric scalar model is shown in Figs. $\ref{fig:full-o0-field}$~-~$\ref{fig:full-o4-field}$ for $N=0~-~4$. We have also computed the exponents for the $N=10$ and $N=100$ cases but omitted to show their field dependence, as it is very small. The leading order of the DE, the local potential approximation (LPA) is exact for $O(N \to \infty)$. The anomalous dimension decreases monotonically at large $N$ values with increasing $N$ and vanishes completely in the limit $N\to \infty$. This means that the derivative expansion has to yield very precise predictions for the exponents for large $N$ values. This is reflected in the fact, that the field dependence is very small at $N=10$ and at $N=100$. We have chosen $N=10$ and $100$ as benchmark points to compare our predictions with those of the large-$N$ expansion. 
We also show the field dependence of the $O(1)$ symmetric model, which should give the critical exponents for $Z_2$ universality class. 
This feature is nicely shown in Fig. \ref{fig:full-o1-field}.
Going back to the Figs. $\ref{fig:full-o0-field}$~-~$\ref{fig:full-o4-field}$, we can clearly see, that the field expansion is very stable at NNLO even when one considers the correction of $M_{WHJ}=1$ compared to $M_{WHJ}=0$. Due to this smoothness of predictions from the field expansion at NNLO, we apply the principle of minimal sensitivity for $M_{WHJ}\geq0$. In order to reduce the amount of computation, we have only looked for a PMS solution for the anomalous dimension and accepted the corresponding parameter value as the optimal $\alpha^{opt}$. We have found that $\alpha^{opt}$ depends weakly both on the truncation $M_{WHJ}$ and on the $O(N)$ model considered.
We have found at NNLO with truncation $M_{WHJ}=4$ for the regulator (\ref{eq:regulator})  that $\alpha^{opt}=0.340(10)$ while for (\ref{eq:exp-regulator}) we have obtained $\alpha^{opt}=0.87(1)$. The uncertainty corresponds to the dependence of $\alpha^{opt}$ on the specific $O(N)$ model considered. For instance, we obtained $\alpha^{opt}=0.337$ for the regulator (\ref{eq:regulator}) in case of the $O(0)$ model with truncation $M_{WHJ}=4$ and $\alpha^{opt}=0.344$ in case of the $O(3)$ model in the same setting.
We attempt to find the limiting value of the optimized exponents
corresponding to $M_{WHJ}=0,\ldots,4$
in the range $N=0~-~4$ for $M_{WHJ} \to \infty$ in the same fashion as we did for the $Z_2$ symmetric model (see Fig. \ref{fig:full-o1-opt}). 
We have also checked the stability of the predictions from (\ref{eq:resumtaylor}). We have computed the extrapolated values of the critical exponents from (\ref{eq:resumtaylor}) by fitting those to the PMS optimized exponents corresponding to $M_{WHJ}=0,\ldots,4$ and $M_{WHJ}=0,\ldots,3$. In the latter case we also incorporated an assumption for the fit. Namely, whether we expect (from the trend of the polynomial expansion) $X(M_{WHJ}=4)$ to be greater or smaller than $X(M_{WHJ}=3)$. As a result we obtained that the difference between the predictions obtained from the fits are $2-3$ times smaller than the difference between the raw values corresponding to $M_{WHJ}=4$ and $M_{WHJ}=3$. This is the method to obtain the uncertainty $\Delta_{poly}\overline{X}^{(4)}$ detailed in App.~\ref{app:error-estimate}.
As for $N=10$ and $N=100$ the fluctuation of the exponents is very small with varying $M_{WHJ}$. In these instances we consider our final predictions corresponding to $M_{WHJ}=3$ and $M_{WHJ}=2$ respectively, with PMS optimization applied. 
\begin{table}[hbt!]
    \centering
\begin{tabular}{ |c|c||c c c|c| } 
\hline
$N$ & Order of DE & $\nu$ & $\eta$ & $\omega$\\
\hline
\hline
\multirow{3}{2cm}{~~~0} 
& LPA & $0.5924(3)$ & $0$ & $0.656(2)$\\ 
& NLO & $0.588(1)$ & $0.038(9)$ & $0.95(8)$\\ 
& NNLO & $0.5876(1)$ & $0.030(3)$ & $0.894(16)$\\ 
\hline
\hline
\multirow{3}{4em}{1} 
& LPA & $0.6504(7)$ & $0$ & $0.654(1)$\\ 
& NLO & $0.628(6)$ & $0.045(11)$ & $0.85(5)$\\ 
& NNLO & $0.630(1)$ & $0.035(3)$ & $0.829(6)$\\ 
\hline
\hline
\multirow{3}{4em}{2} 
& LPA & $0.7098(10)$ & $0$ & $0.672(1)$\\ 
& NLO & $0.667(10)$ & $0.047(13)$ & $0.79(3)$\\ 
& NNLO & $0.673(2)$ & $0.036(3)$ & $0.784(8)$\\ 
\hline
\hline
\multirow{3}{4em}{3} 
& LPA & $0.7629(12)$ & $0$ & $0.702(1)$\\ 
& NLO & $0.705(15)$ & $0.047(13)$ & $0.75(3)$\\ 
& NNLO & $0.713(3)$ & $0.036(3)$ & $0.765(3)$\\ 
\hline
\hline
\multirow{3}{4em}{4} 
& LPA & $0.8060(12)$ & $0$ & $0.737(2)$\\
& NLO & $0.741(20)$ & $0.045(13)$ & $0.73(3)$\\ 
& NNLO & $0.749(3)$ & $0.034(3)$ & $0.763(9)$\\ 
\hline
\hline
\multirow{3}{4em}{10} 
& LPA & $0.9193(5)$ & $0$ & $0.874(2)$\\
& NLO & $0.878(10)$ & $0.027(7)$ & $0.79(2)$\\ 
& NNLO & $0.877(1)$ & $0.022(2)$ & $0.810(7)$\\ 
\hline
\hline
\multirow{3}{4em}{100} 
& LPA & $0.9925(1)$ & $0$ & $0.9881(2)$\\
& NLO & $0.989(1)$ & $0.0030(7)$ & $0.978(3)$\\ 
& NNLO & $0.9888(3)$ & $0.00264(8)$ & $0.9780(6)$\\ 
\hline
\end{tabular}
    \caption{The main findings of this work. Our predictions for the critical exponents $\nu$, $\eta$ and $\omega$ at the LPA, NLO and NNLO of the DE for the $O(N)$ symmetric models in $d=3$ Euclidean dimensions. These values are the average of the PMS optimized predictions, computed from the $\Theta$-regulator (\ref{eq:regulator}) and the exponential regulator (\ref{eq:exp-regulator}) and the deviation from the average is one source of the uncertainties.
    The other source of uncertainty correspond to the polynomial truncation of the scale dependent functions.}
    \label{tab:on-data}
\end{table}
\begin{table}[hbt!]
    \centering
\begin{tabular}{ |c|c||c c c|c| } 
\hline
$N$ & Method & $\nu$ & $\eta$ & $\omega$\\
\hline
\hline
%\rowcolor{lightgray}
\multirow{7}{0.5cm}{0} 
& this work & $0.5875(1)$ & $0.031(3)$ & $0.903(16)$\\
& $\partial^4$, raw & $0.5875$ & $0.0292$ & $0.901$\\ 
& $\partial^4$, improved & $0.5876(2)$ & $0.0312(9)$ & $0.901(24)$\\ 
& MC~\cite{o0-MC1,o0-MC2} & $0.58759700(40)$ & $0.0310434(30)$ & $0.899(14)$\\
& six-loop PT & $0.5882(11)$ & $0.0284(25)$ & $0.812(16)$\\
& $\epsilon^6$, $\epsilon$-exp. & $0.5874(3)$ & $0.0310(7)$ & $0.841(13)$\\
& CB~\cite{o0-CB} & $0.5876(12)$ & $0.0282(4)$ &  \\
\hline
\hline
\multirow{7}{0.5cm}{2} 
& this work & $0.672(2)$ & $0.038(3)$ & $0.784(8)$\\
& $\partial^4$, raw & $0.6732$ & $0.0350$ & $0.793$\\
& $\partial^4$, improved & $0.6716(6)$ & $0.0380(13)$ & $0.791(8)$\\
& MC~\cite{MC-XY} & $0.67169(7)$ & $0.03810(8)$ & $0.789(4)$\\
& six-loop PT & $0.6703(15)$ & $0.0354(25)$ & $0.789(11)$\\
& $\epsilon^6$, $\epsilon$-exp. & $0.6690(10)$ & $0.0380(6)$ & $0.804(3)$\\
& CB~\cite{o2-CB} & $0.6718(1)$ & $0.03818(4)$ & $0.794(8)$ \\
\hline
\hline
\multirow{7}{0.5cm}{3} 
& this work & $0.712(3)$ & $0.038(3)$ & $0.765(3)$\\
& $\partial^4$, raw & $0.7136$ & $0.0347$ & $0.773$\\
& $\partial^4$, improved & $0.7114(9)$ & $0.0376(13)$ & $0.769(11)$\\
& MC~\cite{o3-MC,o3-MC2} & $0.7116(10)$ & $0.0378(3)$ & $0.773$\\
& six-loop PT & $0.7073(35)$ & $0.0355(25)$ & $0.782(13)$\\
& $\epsilon^6$, $\epsilon$-exp. & $0.7059(20)$ & $0.0378(5)$ & $0.795(7)$\\
& CB~\cite{o3-CB1,o3-CB2} & $0.7120(23)$ & $0.0385(13)$ & $0.791(22)$\\
\hline
\hline
\multirow{7}{0.5cm}{4} 
& this work & $0.748(3)$ & $0.036(3)$ & $0.763(9)$\\
& $\partial^4$, raw & $0.7500$ & $0.0332$ & $0.765$\\
& $\partial^4$, improved & $0.7478(9)$ & $0.0360(12)$ & $0.761(12)$\\
& MC~\cite{o4-MC1,o3-MC2} & $0.7477(8)$ & $0.0360(4)$ & $0.765$\\
& six-loop PT & $0.741(6)$ & $0.0350(45)$ & $0.774(20)$\\
& $\epsilon^6$, $\epsilon$-exp. & $0.7397(35)$ & $0.0366(4)$ & $0.794(9)$\\
& CB~\cite{o4-CB1,o3-CB2} & $0.7472(87)$ & $0.0378(32)$ & $0.817(30)$ \\
\hline
\hline
\multirow{4}{0.5cm}{10}
& this work & $0.877(1)$ & $0.023(2)$ & $0.805(7)$\\
& $\partial^4$, raw & $0.8771$ & $0.0218$ & $0.808$\\
& $\partial^4$, improved & $0.8776(10)$ & $0.0231(6)$ & $0.807(7)$\\
& large-$N$ & $0.87(2)$ & $0.023(2)$ & $0.77(1)$\\
\hline
\hline
\multirow{4}{0.5cm}{100}
& this work & $0.9887(3)$ & $0.00267(8)$ & $0.9780(6)$\\
& $\partial^4$, raw & $0.98877$ & $0.00260$ & $0.977$\\
& $\partial^4$, improved & $0.9888(2)$ & $0.00268(4)$ & $0.9770(8)$\\
& large-$N$ & $0.9890(2)$ & $0.002681(1)$ & $0.9782(2)$\\
\hline
\end{tabular}
    \caption{Critical exponents of the $O(N)$ symmetric scalar model in $d=3$ Euclidean dimensions for several $N$ values with different methods: our improved predictions using Wynn's epsilon algorithm, the
    DE at NNLO ($\partial^4$) without field expansion with raw (computed with the exponential regulator) and improved values \cite{full-on}, Monte-Carlo simulations, six-loop perturbation theory at fixed $d=3$ \cite{Guida_1998}, $d=4-\epsilon$ expansion at $\epsilon^6$ \cite{eps-expand}, the conformal bootstrap method and the large-$N$ expansion \cite{large-N,o10-largeN1,o10-largeN2}. \label{tab:on-refsum}
    }
\end{table}

Considering the above discussed details, our predictions for $O(N)$ critical exponents at fixed orders of the DE are summarized in Tab.~\ref{tab:on-data}. Our findings at the level of LPA correspond to the exponents computed at $M_U=8$ with the method detailed at the end of Sec.~\ref{sec:regulator}. We obtain that the optimal value $\alpha^{opt}_{LPA}$ of the parameter $\alpha$ is $0.9$ for (\ref{eq:regulator}) and $5$ for (\ref{eq:exp-regulator}) at LPA. Going further, our NLO findings are computed at $M_U=8$ and $M_{ZY}=5$ with $\alpha^{opt}_{NLO}=0.4$ and $1.4$ for (\ref{eq:regulator}) and (\ref{eq:exp-regulator}) respectively. The results for the NNLO level results are discussed above. 
The method we used to compute the central values and the uncertainties are detailed in App.~\ref{app:error-estimate}.

\section{Brief summary of the $O(N)$ critical exponents from various methods}
\label{sect:on-sum}
The $O(N)$ symmetric scalar model was first introduced as the $n$-vector model as a generalization of some physically relevant models \cite{on-nvect} in $d$ Euclidean dimensions. The $N=0$ case describes the self avoiding walk \cite{o0-first,o0-second}. It is also noteworthy, that the $O(0)$ model probably does not have a Minkowskian counterpart, because in case the Euclidean dimension $d$ and $N$ are not positive integers the unitarity of the corresponding Minkowskian model is lost or at least highly nontrivial. 
At the level of the $n$-vector model, $O(1)$ model describes the Ising universality class. In the ERG however the $Z_2$ and $O(N)$ symmetric models at $N=1$ seem to be different because of the different content of scale dependent functions and the appearance of an additional, massless excitation in the $O(N)$ model. The two models are equivalent however. The flow equations for the $O(N)$ model in the limit of $N \to 1$ are regular. Furthermore the contribution of the Goldstone modes in the flow equations vanish for $N=1$, and the extra flow equations decouple from those which have direct interpretation in terms of the $Z_2$ symmetric model.
The $O(2)$ model is more commonly known as the $XY$-model, which is used to describe the phase transition in the superfluid helium-4. The $O(3)$ model is also known as the Heisenberg model for ferromagnetism. Lastly but not the least, the $O(4)$ model can be considered as a toy model for the standard model's Higgs sector, but also applicable to chiral phase transitions. 

Some of the most precise computations of the $O(N)$ critical exponents in $d=3$ Euclidean dimensions are summarized in Tab.~\ref{tab:on-refsum}. Comparing these with our findings, 'this work' entry in the same table, one can see that the central values are in excellent agreement. The improved results of Ref.~\cite{full-on} take advantage of the convergence of the DE as well as the alternating behavior of the corrections from the successive orders of the DE.
In contrast our improvement, the Wynn epsilon algorithm detailed in Sec.~\ref{sec:wynn}, is a robust series acceleration method applicable to any alternating sequence. 

\section{Conclusion}

We have computed the critical exponents for the $Z_2$ and $O(N)$ symmetric scalar models in $d=3$ Euclidean dimensions. We have employed the exact renormalization group equation for the effective average action. We have used the derivative expansion at NNLO (or $\partial^4$-order) and calculated the $\beta$-functions for the scale dependent functions, shown in (\ref{eq:z2-nnlo}) for the $Z_2$ and in (\ref{eq:fullon}) for the $O(N)$ symmetric models. In order to locate the Wilson-Fisher fixed point which is the nontrivial fixed point solution of the beta-functions, we have expanded the scale dependent functions in powers of the field. We interpret the scale dependent coefficients $f_n(k)$ from the Taylor expansion as effective coupling strengths for the interaction vertices of the field they multiply. We have located the fixed point in the theory space spanned by the (canonical mass) dimensionless couplings, with truncated Taylor series of the scale dependent functions. Our main findings for the $Z_2$ symmetric model shown in Tab.~\ref{tab:o1-summary} are in agreement with predictions obtained using other methods. We have used the $Z_2$ symmetric model as a testing ground for the correctness of our \texttt{Mathematica} code. We then generalized this code for the $O(N)$ symmetric model and computed the critical exponents for some relevant $N$ values. We have tested the $O(N)$ \texttt{Mathematica} code for the $N=1$, 10 and 100 cases. 
The first benchmark point $N=1$ is chosen, because it should reproduce the Ising critical exponents as the $O(1)$ and $Z_2$ symmetric models are equivalent as discussed in Sect.~\ref{sect:on-sum}.
We chose $N=10,~100$ to be second and third benchmark points, because the effect of the derivative expansion is diminished with $N \to \infty$, hence it can give very accurate results for large $N$ values. Our main findings are summarized in Tab.~\ref{tab:on-data}. 
A great advantage of the computations employed in this work is that they require noticeably less computer time than most of the other methods. 
For our highest employed polynomial truncation both for the $Z_2$ and $O(N)$ symmetric models, the location of the Wilson-Fisher fixed point roughly takes $1~-~2$ hours, while computing the Jacobian matrix at the fixed point takes an additional hour on a single desktop PC. 

In a recent paper \cite{full-on} the authors have performed similar computations with the ERG. The differences are that (i) we have not truncated our formulas in the momenta (denoted here with $Q_i$); (ii) we have employed Taylor expansion for the scale dependent functions in powers of the field instead of shooting for a solution for the complete scale dependent functions; (iii) we have computed the exponents with the regulator (\ref{eq:regulator}), which is the simplest regulator at NNLO. Although this $\Theta$-regulator is argued to perform poorly in \cite{PhysRevLett.123.240604}, we have found that it yields excellent predictions for the exponents in the models studied here. We also provide improved predictions using Wynn's espilon algorithm on our predictions of the DE, yielding central values which are in excellent agreement with other precise methods used to compute critical exponents.

We also produces the subleading scaling corrections $\omega_i$ (from the eigenvalue spectrum $-1/\nu<\omega<\omega_1<\omega_2<\ldots$ of the Jacobian of the beta-functions) as a byproduct of computing the exponents $\nu$ and $\omega$. The expansion of the scale dependent functions in powers of the field is also applicable to explore the phase structure of a model and the RG running of its couplings. The derivative expansion can also be improved to N${^3}$LO (or $\partial^6$-order) with some effort for the $O(N)$ symmetric models, which would provide more precise exponent values for many cases of $N$.

\section*{Acknowledgments}
The author would like to thank Z. Trócsányi for the careful reading of the manuscript.

\bibliography{PPMCISU}

\providecommand{\noopsort}[1]{}\providecommand{\singleletter}[1]{#1}%
\begin{thebibliography}{40}%
\makeatletter
\providecommand \@ifxundefined [1]{%
 \@ifx{#1\undefined}
}%
\providecommand \@ifnum [1]{%
 \ifnum #1\expandafter \@firstoftwo
 \else \expandafter \@secondoftwo
 \fi
}%
\providecommand \@ifx [1]{%
 \ifx #1\expandafter \@firstoftwo
 \else \expandafter \@secondoftwo
 \fi
}%
\providecommand \natexlab [1]{#1}%
\providecommand \enquote  [1]{``#1''}%
\providecommand \bibnamefont  [1]{#1}%
\providecommand \bibfnamefont [1]{#1}%
\providecommand \citenamefont [1]{#1}%
\providecommand \href@noop [0]{\@secondoftwo}%
\providecommand \href [0]{\begingroup \@sanitize@url \@href}%
\providecommand \@href[1]{\@@startlink{#1}\@@href}%
\providecommand \@@href[1]{\endgroup#1\@@endlink}%
\providecommand \@sanitize@url [0]{\catcode `\\12\catcode `\$12\catcode
  `\&12\catcode `\#12\catcode `\^12\catcode `\_12\catcode `\%12\relax}%
\providecommand \@@startlink[1]{}%
\providecommand \@@endlink[0]{}%
\providecommand \url  [0]{\begingroup\@sanitize@url \@url }%
\providecommand \@url [1]{\endgroup\@href {#1}{\urlprefix }}%
\providecommand \urlprefix  [0]{URL }%
\providecommand \Eprint [0]{\href }%
\providecommand \doibase [0]{https://doi.org/}%
\providecommand \selectlanguage [0]{\@gobble}%
\providecommand \bibinfo  [0]{\@secondoftwo}%
\providecommand \bibfield  [0]{\@secondoftwo}%
\providecommand \translation [1]{[#1]}%
\providecommand \BibitemOpen [0]{}%
\providecommand \bibitemStop [0]{}%
\providecommand \bibitemNoStop [0]{.\EOS\space}%
\providecommand \EOS [0]{\spacefactor3000\relax}%
\providecommand \BibitemShut  [1]{\csname bibitem#1\endcsname}%
\let\auto@bib@innerbib\@empty
%</preamble>
\bibitem [{\citenamefont {Wetterich}(1993)}]{WETTERICH199390}%
  \BibitemOpen
  \bibfield  {author} {\bibinfo {author} {\bibfnamefont {C.}~\bibnamefont
  {Wetterich}},\ }\bibfield  {title} {\bibinfo {title} {Exact evolution
  equation for the effective potential},\ }\href
  {https://doi.org/https://doi.org/10.1016/0370-2693(93)90726-X} {\bibfield
  {journal} {\bibinfo  {journal} {Physics Letters B}\ }\textbf {\bibinfo
  {volume} {301}},\ \bibinfo {pages} {90 } (\bibinfo {year}
  {1993})}\BibitemShut {NoStop}%
\bibitem [{\citenamefont {Wilson}\ and\ \citenamefont
  {Kogut}(1974)}]{WILSON197475}%
  \BibitemOpen
  \bibfield  {author} {\bibinfo {author} {\bibfnamefont {K.~G.}\ \bibnamefont
  {Wilson}}\ and\ \bibinfo {author} {\bibfnamefont {J.}~\bibnamefont {Kogut}},\
  }\bibfield  {title} {\bibinfo {title} {The renormalization group and the
  $\epsilon$ expansion},\ }\href
  {https://doi.org/https://doi.org/10.1016/0370-1573(74)90023-4} {\bibfield
  {journal} {\bibinfo  {journal} {Physics Reports}\ }\textbf {\bibinfo {volume}
  {12}},\ \bibinfo {pages} {75 } (\bibinfo {year} {1974})}\BibitemShut
  {NoStop}%
\bibitem [{\citenamefont {Hasenbusch}(2010)}]{MC-ising}%
  \BibitemOpen
  \bibfield  {author} {\bibinfo {author} {\bibfnamefont {M.}~\bibnamefont
  {Hasenbusch}},\ }\bibfield  {title} {\bibinfo {title} {Finite size scaling
  study of lattice models in the three-dimensional ising universality class},\
  }\href {https://doi.org/10.1103/PhysRevB.82.174433} {\bibfield  {journal}
  {\bibinfo  {journal} {Phys. Rev. B}\ }\textbf {\bibinfo {volume} {82}},\
  \bibinfo {pages} {174433} (\bibinfo {year} {2010})}\BibitemShut {NoStop}%
\bibitem [{\citenamefont {Hasenbusch}(2019)}]{MC-XY}%
  \BibitemOpen
  \bibfield  {author} {\bibinfo {author} {\bibfnamefont {M.}~\bibnamefont
  {Hasenbusch}},\ }\bibfield  {title} {\bibinfo {title} {Monte carlo study of
  an improved clock model in three dimensions},\ }\href
  {https://doi.org/10.1103/PhysRevB.100.224517} {\bibfield  {journal} {\bibinfo
   {journal} {Phys. Rev. B}\ }\textbf {\bibinfo {volume} {100}},\ \bibinfo
  {pages} {224517} (\bibinfo {year} {2019})}\BibitemShut {NoStop}%
\bibitem [{\citenamefont {Guida}\ and\ \citenamefont
  {Zinn-Justin}(1998)}]{Guida_1998}%
  \BibitemOpen
  \bibfield  {author} {\bibinfo {author} {\bibfnamefont {R.}~\bibnamefont
  {Guida}}\ and\ \bibinfo {author} {\bibfnamefont {J.}~\bibnamefont
  {Zinn-Justin}},\ }\bibfield  {title} {\bibinfo {title} {Critical exponents of
  {theN}-vector model},\ }\href {https://doi.org/10.1088/0305-4470/31/40/006}
  {\bibfield  {journal} {\bibinfo  {journal} {Journal of Physics A:
  Mathematical and General}\ }\textbf {\bibinfo {volume} {31}},\ \bibinfo
  {pages} {8103} (\bibinfo {year} {1998})}\BibitemShut {NoStop}%
\bibitem [{\citenamefont {Schnetz}(2018)}]{Loop-ising}%
  \BibitemOpen
  \bibfield  {author} {\bibinfo {author} {\bibfnamefont {O.}~\bibnamefont
  {Schnetz}},\ }\bibfield  {title} {\bibinfo {title} {Numbers and functions in
  quantum field theory},\ }\href {https://doi.org/10.1103/PhysRevD.97.085018}
  {\bibfield  {journal} {\bibinfo  {journal} {Phys. Rev. D}\ }\textbf {\bibinfo
  {volume} {97}},\ \bibinfo {pages} {085018} (\bibinfo {year}
  {2018})}\BibitemShut {NoStop}%
\bibitem [{\citenamefont {Kompaniets}\ and\ \citenamefont
  {Panzer}(2017)}]{eps-expand}%
  \BibitemOpen
  \bibfield  {author} {\bibinfo {author} {\bibfnamefont {M.~V.}\ \bibnamefont
  {Kompaniets}}\ and\ \bibinfo {author} {\bibfnamefont {E.}~\bibnamefont
  {Panzer}},\ }\bibfield  {title} {\bibinfo {title} {Minimally subtracted
  six-loop renormalization of $o(n)$-symmetric ${\ensuremath{\phi}}^{4}$ theory
  and critical exponents},\ }\href {https://doi.org/10.1103/PhysRevD.96.036016}
  {\bibfield  {journal} {\bibinfo  {journal} {Phys. Rev. D}\ }\textbf {\bibinfo
  {volume} {96}},\ \bibinfo {pages} {036016} (\bibinfo {year}
  {2017})}\BibitemShut {NoStop}%
\bibitem [{\citenamefont {El-Showk}\ \emph {et~al.}(2012)\citenamefont
  {El-Showk}, \citenamefont {Paulos}, \citenamefont {Poland}, \citenamefont
  {Rychkov}, \citenamefont {Simmons-Duffin},\ and\ \citenamefont
  {Vichi}}]{bootstrap-ising}%
  \BibitemOpen
  \bibfield  {author} {\bibinfo {author} {\bibfnamefont {S.}~\bibnamefont
  {El-Showk}}, \bibinfo {author} {\bibfnamefont {M.~F.}\ \bibnamefont
  {Paulos}}, \bibinfo {author} {\bibfnamefont {D.}~\bibnamefont {Poland}},
  \bibinfo {author} {\bibfnamefont {S.}~\bibnamefont {Rychkov}}, \bibinfo
  {author} {\bibfnamefont {D.}~\bibnamefont {Simmons-Duffin}},\ and\ \bibinfo
  {author} {\bibfnamefont {A.}~\bibnamefont {Vichi}},\ }\bibfield  {title}
  {\bibinfo {title} {Solving the 3d ising model with the conformal bootstrap},\
  }\href {https://doi.org/10.1103/PhysRevD.86.025022} {\bibfield  {journal}
  {\bibinfo  {journal} {Phys. Rev. D}\ }\textbf {\bibinfo {volume} {86}},\
  \bibinfo {pages} {025022} (\bibinfo {year} {2012})}\BibitemShut {NoStop}%
\bibitem [{\citenamefont {Simmons-Duffin}(2015)}]{CB-cost}%
  \BibitemOpen
  \bibfield  {author} {\bibinfo {author} {\bibfnamefont {D.}~\bibnamefont
  {Simmons-Duffin}},\ }\bibfield  {title} {\bibinfo {title} {A semidefinite
  program solver for the conformal bootstrap},\ }\href
  {https://doi.org/10.1007/JHEP06(2015)174} {\bibfield  {journal} {\bibinfo
  {journal} {Journal of High Energy Physics}\ }\textbf {\bibinfo {volume}
  {2015}},\ \bibinfo {pages} {174} (\bibinfo {year} {2015})}\BibitemShut
  {NoStop}%
\bibitem [{\citenamefont {Balog}\ \emph {et~al.}(2019)\citenamefont {Balog},
  \citenamefont {Chat\'e}, \citenamefont {Delamotte}, \citenamefont
  {Marohni\ifmmode~\acute{c}\else \'{c}\fi{}},\ and\ \citenamefont
  {Wschebor}}]{PhysRevLett.123.240604}%
  \BibitemOpen
  \bibfield  {author} {\bibinfo {author} {\bibfnamefont {I.}~\bibnamefont
  {Balog}}, \bibinfo {author} {\bibfnamefont {H.}~\bibnamefont {Chat\'e}},
  \bibinfo {author} {\bibfnamefont {B.}~\bibnamefont {Delamotte}}, \bibinfo
  {author} {\bibfnamefont {M.}~\bibnamefont {Marohni\ifmmode~\acute{c}\else
  \'{c}\fi{}}},\ and\ \bibinfo {author} {\bibfnamefont {N.}~\bibnamefont
  {Wschebor}},\ }\bibfield  {title} {\bibinfo {title} {Convergence of
  nonperturbative approximations to the renormalization group},\ }\href
  {https://doi.org/10.1103/PhysRevLett.123.240604} {\bibfield  {journal}
  {\bibinfo  {journal} {Phys. Rev. Lett.}\ }\textbf {\bibinfo {volume} {123}},\
  \bibinfo {pages} {240604} (\bibinfo {year} {2019})}\BibitemShut {NoStop}%
\bibitem [{\citenamefont {De~Polsi}\ \emph {et~al.}(2020)\citenamefont
  {De~Polsi}, \citenamefont {Balog}, \citenamefont {Tissier},\ and\
  \citenamefont {Wschebor}}]{full-on}%
  \BibitemOpen
  \bibfield  {author} {\bibinfo {author} {\bibfnamefont {G.}~\bibnamefont
  {De~Polsi}}, \bibinfo {author} {\bibfnamefont {I.}~\bibnamefont {Balog}},
  \bibinfo {author} {\bibfnamefont {M.}~\bibnamefont {Tissier}},\ and\ \bibinfo
  {author} {\bibfnamefont {N.}~\bibnamefont {Wschebor}},\ }\bibfield  {title}
  {\bibinfo {title} {Precision calculation of critical exponents in the o(n)
  universality classes with the nonperturbative renormalization group},\ }\href
  {https://doi.org/10.1103/PhysRevE.101.042113} {\bibfield  {journal} {\bibinfo
   {journal} {Phys. Rev. E}\ }\textbf {\bibinfo {volume} {101}},\ \bibinfo
  {pages} {042113} (\bibinfo {year} {2020})}\BibitemShut {NoStop}%
\bibitem [{\citenamefont {Litim}(2001)}]{Litim2001}%
  \BibitemOpen
  \bibfield  {author} {\bibinfo {author} {\bibfnamefont {D.~F.}\ \bibnamefont
  {Litim}},\ }\bibfield  {title} {\bibinfo {title} {Optimized renormalization
  group flows},\ }\href {https://doi.org/10.1103/PhysRevD.64.105007} {\bibfield
   {journal} {\bibinfo  {journal} {Phys. Rev. D}\ }\textbf {\bibinfo {volume}
  {64}},\ \bibinfo {pages} {105007} (\bibinfo {year} {2001})}\BibitemShut
  {NoStop}%
\bibitem [{\citenamefont {Canet}\ \emph
  {et~al.}(2003{\natexlab{a}})\citenamefont {Canet}, \citenamefont {Delamotte},
  \citenamefont {Mouhanna},\ and\ \citenamefont {Vidal}}]{dela2003}%
  \BibitemOpen
  \bibfield  {author} {\bibinfo {author} {\bibfnamefont {L.}~\bibnamefont
  {Canet}}, \bibinfo {author} {\bibfnamefont {B.}~\bibnamefont {Delamotte}},
  \bibinfo {author} {\bibfnamefont {D.}~\bibnamefont {Mouhanna}},\ and\
  \bibinfo {author} {\bibfnamefont {J.}~\bibnamefont {Vidal}},\ }\bibfield
  {title} {\bibinfo {title} {Nonperturbative renormalization group approach to
  the ising model: A derivative expansion at order
  ${\ensuremath{\partial}}^{4}$},\ }\href
  {https://doi.org/10.1103/PhysRevB.68.064421} {\bibfield  {journal} {\bibinfo
  {journal} {Phys. Rev. B}\ }\textbf {\bibinfo {volume} {68}},\ \bibinfo
  {pages} {064421} (\bibinfo {year} {2003}{\natexlab{a}})}\BibitemShut
  {NoStop}%
\bibitem [{\citenamefont {D'Attanasio}\ and\ \citenamefont
  {Morris}(1997)}]{largeN-refree}%
  \BibitemOpen
  \bibfield  {author} {\bibinfo {author} {\bibfnamefont {M.}~\bibnamefont
  {D'Attanasio}}\ and\ \bibinfo {author} {\bibfnamefont {T.~R.}\ \bibnamefont
  {Morris}},\ }\bibfield  {title} {\bibinfo {title} {{Large N and the
  renormalization group}},\ }\href
  {https://doi.org/10.1016/S0370-2693(97)00866-6} {\bibfield  {journal}
  {\bibinfo  {journal} {Phys. Lett. B}\ }\textbf {\bibinfo {volume} {409}},\
  \bibinfo {pages} {363} (\bibinfo {year} {1997})},\ \Eprint
  {https://arxiv.org/abs/hep-th/9704094} {arXiv:hep-th/9704094} \BibitemShut
  {NoStop}%
\bibitem [{\citenamefont {Stevenson}(1981)}]{PMS-opt}%
  \BibitemOpen
  \bibfield  {author} {\bibinfo {author} {\bibfnamefont {P.~M.}\ \bibnamefont
  {Stevenson}},\ }\bibfield  {title} {\bibinfo {title} {Optimized perturbation
  theory},\ }\href {https://doi.org/10.1103/PhysRevD.23.2916} {\bibfield
  {journal} {\bibinfo  {journal} {Phys. Rev. D}\ }\textbf {\bibinfo {volume}
  {23}},\ \bibinfo {pages} {2916} (\bibinfo {year} {1981})}\BibitemShut
  {NoStop}%
\bibitem [{\citenamefont {Canet}\ \emph
  {et~al.}(2003{\natexlab{b}})\citenamefont {Canet}, \citenamefont {Delamotte},
  \citenamefont {Mouhanna},\ and\ \citenamefont {Vidal}}]{dela-nlo}%
  \BibitemOpen
  \bibfield  {author} {\bibinfo {author} {\bibfnamefont {L.}~\bibnamefont
  {Canet}}, \bibinfo {author} {\bibfnamefont {B.}~\bibnamefont {Delamotte}},
  \bibinfo {author} {\bibfnamefont {D.}~\bibnamefont {Mouhanna}},\ and\
  \bibinfo {author} {\bibfnamefont {J.}~\bibnamefont {Vidal}},\ }\bibfield
  {title} {\bibinfo {title} {Optimization of the derivative expansion in the
  nonperturbative renormalization group},\ }\href
  {https://doi.org/10.1103/PhysRevD.67.065004} {\bibfield  {journal} {\bibinfo
  {journal} {Phys. Rev. D}\ }\textbf {\bibinfo {volume} {67}},\ \bibinfo
  {pages} {065004} (\bibinfo {year} {2003}{\natexlab{b}})}\BibitemShut
  {NoStop}%
\bibitem [{\citenamefont {Aoki}\ \emph {et~al.}(1998)\citenamefont {Aoki},
  \citenamefont {Morikawa}, \citenamefont {Souma}, \citenamefont {Sumi},\ and\
  \citenamefont {Terao}}]{poly-stability}%
  \BibitemOpen
  \bibfield  {author} {\bibinfo {author} {\bibfnamefont {K.-I.}\ \bibnamefont
  {Aoki}}, \bibinfo {author} {\bibfnamefont {K.}~\bibnamefont {Morikawa}},
  \bibinfo {author} {\bibfnamefont {W.}~\bibnamefont {Souma}}, \bibinfo
  {author} {\bibfnamefont {J.-I.}\ \bibnamefont {Sumi}},\ and\ \bibinfo
  {author} {\bibfnamefont {H.}~\bibnamefont {Terao}},\ }\bibfield  {title}
  {\bibinfo {title} {{Rapidly Converging Truncation Scheme of the Exact
  Renormalization Group}},\ }\href {https://doi.org/10.1143/PTP.99.451}
  {\bibfield  {journal} {\bibinfo  {journal} {Progress of Theoretical Physics}\
  }\textbf {\bibinfo {volume} {99}},\ \bibinfo {pages} {451} (\bibinfo {year}
  {1998})},\ \Eprint
  {https://arxiv.org/abs/https://academic.oup.com/ptp/article-pdf/99/3/451/5222556/99-3-451.pdf}
  {https://academic.oup.com/ptp/article-pdf/99/3/451/5222556/99-3-451.pdf}
  \BibitemShut {NoStop}%
\bibitem [{\citenamefont {Litim}(2002)}]{LITIM2002128}%
  \BibitemOpen
  \bibfield  {author} {\bibinfo {author} {\bibfnamefont {D.~F.}\ \bibnamefont
  {Litim}},\ }\bibfield  {title} {\bibinfo {title} {Critical exponents from
  optimised renormalisation group flows},\ }\href
  {https://doi.org/https://doi.org/10.1016/S0550-3213(02)00186-4} {\bibfield
  {journal} {\bibinfo  {journal} {Nuclear Physics B}\ }\textbf {\bibinfo
  {volume} {631}},\ \bibinfo {pages} {128 } (\bibinfo {year}
  {2002})}\BibitemShut {NoStop}%
\bibitem [{\citenamefont {Peli}\ \emph {et~al.}(2018)\citenamefont {Peli},
  \citenamefont {Nagy},\ and\ \citenamefont {Sailer}}]{polyexp-bound}%
  \BibitemOpen
  \bibfield  {author} {\bibinfo {author} {\bibfnamefont {Z.}~\bibnamefont
  {Peli}}, \bibinfo {author} {\bibfnamefont {S.}~\bibnamefont {Nagy}},\ and\
  \bibinfo {author} {\bibfnamefont {K.}~\bibnamefont {Sailer}},\ }\bibfield
  {title} {\bibinfo {title} {{Effect of the quartic gradient terms on the
  critical exponents of the Wilson-Fisher fixed point in $O(N)$ models}},\
  }\href {https://doi.org/10.1140/epja/i2018-12385-9} {\bibfield  {journal}
  {\bibinfo  {journal} {Eur. Phys. J. A}\ }\textbf {\bibinfo {volume} {54}},\
  \bibinfo {pages} {20} (\bibinfo {year} {2018})},\ \Eprint
  {https://arxiv.org/abs/1704.07087} {arXiv:1704.07087 [hep-th]} \BibitemShut
  {NoStop}%
\bibitem [{\citenamefont {Wynn}(1956)}]{wynn1}%
  \BibitemOpen
  \bibfield  {author} {\bibinfo {author} {\bibfnamefont {P.}~\bibnamefont
  {Wynn}},\ }\bibfield  {title} {\bibinfo {title} {On a device for computing
  the em(sn) transformation},\ }\href {http://www.jstor.org/stable/2002183}
  {\bibfield  {journal} {\bibinfo  {journal} {Mathematical Tables and Other
  Aids to Computation}\ }\textbf {\bibinfo {volume} {10}},\ \bibinfo {pages}
  {91} (\bibinfo {year} {1956})}\BibitemShut {NoStop}%
\bibitem [{\citenamefont {Graves-Morris}\ \emph {et~al.}(2000)\citenamefont
  {Graves-Morris}, \citenamefont {Roberts},\ and\ \citenamefont
  {Salam}}]{wynn2}%
  \BibitemOpen
  \bibfield  {author} {\bibinfo {author} {\bibfnamefont {P.}~\bibnamefont
  {Graves-Morris}}, \bibinfo {author} {\bibfnamefont {D.}~\bibnamefont
  {Roberts}},\ and\ \bibinfo {author} {\bibfnamefont {A.}~\bibnamefont
  {Salam}},\ }\bibfield  {title} {\bibinfo {title} {The epsilon algorithm and
  related topics},\ }\href
  {https://doi.org/https://doi.org/10.1016/S0377-0427(00)00355-1} {\bibfield
  {journal} {\bibinfo  {journal} {Journal of Computational and Applied
  Mathematics}\ }\textbf {\bibinfo {volume} {122}},\ \bibinfo {pages} {51 }
  (\bibinfo {year} {2000})},\ \bibinfo {note} {numerical Analysis in the 20th
  Century Vol. II: Interpolation and Extrapolation}\BibitemShut {NoStop}%
\bibitem [{\citenamefont {Tetradis}\ and\ \citenamefont
  {Wetterich}(1994)}]{TETRADIS1994541}%
  \BibitemOpen
  \bibfield  {author} {\bibinfo {author} {\bibfnamefont {N.}~\bibnamefont
  {Tetradis}}\ and\ \bibinfo {author} {\bibfnamefont {C.}~\bibnamefont
  {Wetterich}},\ }\bibfield  {title} {\bibinfo {title} {Critical exponents from
  the effective average action},\ }\href
  {https://doi.org/https://doi.org/10.1016/0550-3213(94)90446-4} {\bibfield
  {journal} {\bibinfo  {journal} {Nuclear Physics B}\ }\textbf {\bibinfo
  {volume} {422}},\ \bibinfo {pages} {541 } (\bibinfo {year}
  {1994})}\BibitemShut {NoStop}%
\bibitem [{\citenamefont {Gersdorff}\ and\ \citenamefont
  {Wetterich}(2001)}]{wette-on}%
  \BibitemOpen
  \bibfield  {author} {\bibinfo {author} {\bibfnamefont {G.~v.}\ \bibnamefont
  {Gersdorff}}\ and\ \bibinfo {author} {\bibfnamefont {C.}~\bibnamefont
  {Wetterich}},\ }\bibfield  {title} {\bibinfo {title} {Nonperturbative
  renormalization flow and essential scaling for the kosterlitz-thouless
  transition},\ }\href {https://doi.org/10.1103/PhysRevB.64.054513} {\bibfield
  {journal} {\bibinfo  {journal} {Phys. Rev. B}\ }\textbf {\bibinfo {volume}
  {64}},\ \bibinfo {pages} {054513} (\bibinfo {year} {2001})}\BibitemShut
  {NoStop}%
\bibitem [{\citenamefont {Clisby}\ and\ \citenamefont
  {D\"unweg}(2016)}]{o0-MC1}%
  \BibitemOpen
  \bibfield  {author} {\bibinfo {author} {\bibfnamefont {N.}~\bibnamefont
  {Clisby}}\ and\ \bibinfo {author} {\bibfnamefont {B.}~\bibnamefont
  {D\"unweg}},\ }\bibfield  {title} {\bibinfo {title} {High-precision estimate
  of the hydrodynamic radius for self-avoiding walks},\ }\href
  {https://doi.org/10.1103/PhysRevE.94.052102} {\bibfield  {journal} {\bibinfo
  {journal} {Phys. Rev. E}\ }\textbf {\bibinfo {volume} {94}},\ \bibinfo
  {pages} {052102} (\bibinfo {year} {2016})}\BibitemShut {NoStop}%
\bibitem [{\citenamefont {Clisby}(2017)}]{o0-MC2}%
  \BibitemOpen
  \bibfield  {author} {\bibinfo {author} {\bibfnamefont {N.}~\bibnamefont
  {Clisby}},\ }\bibfield  {title} {\bibinfo {title} {Scale-free monte carlo
  method for calculating the critical exponent $\gamma$ of self-avoiding
  walks},\ }\href {https://doi.org/10.1088/1751-8121/aa7231} {\bibfield
  {journal} {\bibinfo  {journal} {Journal of Physics A: Mathematical and
  Theoretical}\ }\textbf {\bibinfo {volume} {50}},\ \bibinfo {pages} {264003}
  (\bibinfo {year} {2017})}\BibitemShut {NoStop}%
\bibitem [{\citenamefont {Shimada}\ and\ \citenamefont {Hikami}(2016)}]{o0-CB}%
  \BibitemOpen
  \bibfield  {author} {\bibinfo {author} {\bibfnamefont {H.}~\bibnamefont
  {Shimada}}\ and\ \bibinfo {author} {\bibfnamefont {S.}~\bibnamefont
  {Hikami}},\ }\bibfield  {title} {\bibinfo {title} {Fractal dimensions of
  self-avoiding walks and ising high-temperature graphs in 3d conformal
  bootstrap},\ }\href {https://doi.org/10.1007/s10955-016-1658-x} {\bibfield
  {journal} {\bibinfo  {journal} {Journal of Statistical Physics}\ }\textbf
  {\bibinfo {volume} {165}},\ \bibinfo {pages} {1006} (\bibinfo {year}
  {2016})}\BibitemShut {NoStop}%
\bibitem [{\citenamefont {Chester}\ \emph {et~al.}(2020)\citenamefont
  {Chester}, \citenamefont {Landry}, \citenamefont {Liu}, \citenamefont
  {Poland}, \citenamefont {Simmons-Duffin}, \citenamefont {Su},\ and\
  \citenamefont {Vichi}}]{o2-CB}%
  \BibitemOpen
  \bibfield  {author} {\bibinfo {author} {\bibfnamefont {S.~M.}\ \bibnamefont
  {Chester}}, \bibinfo {author} {\bibfnamefont {W.}~\bibnamefont {Landry}},
  \bibinfo {author} {\bibfnamefont {J.}~\bibnamefont {Liu}}, \bibinfo {author}
  {\bibfnamefont {D.}~\bibnamefont {Poland}}, \bibinfo {author} {\bibfnamefont
  {D.}~\bibnamefont {Simmons-Duffin}}, \bibinfo {author} {\bibfnamefont
  {N.}~\bibnamefont {Su}},\ and\ \bibinfo {author} {\bibfnamefont
  {A.}~\bibnamefont {Vichi}},\ }\bibfield  {title} {\bibinfo {title} {Carving
  out ope space and precise o(2) model critical exponents},\ }\href
  {https://doi.org/10.1007/JHEP06(2020)142} {\bibfield  {journal} {\bibinfo
  {journal} {Journal of High Energy Physics}\ }\textbf {\bibinfo {volume}
  {2020}},\ \bibinfo {pages} {142} (\bibinfo {year} {2020})}\BibitemShut
  {NoStop}%
\bibitem [{\citenamefont {Hasenbusch}\ and\ \citenamefont
  {Vicari}(2011)}]{o3-MC}%
  \BibitemOpen
  \bibfield  {author} {\bibinfo {author} {\bibfnamefont {M.}~\bibnamefont
  {Hasenbusch}}\ and\ \bibinfo {author} {\bibfnamefont {E.}~\bibnamefont
  {Vicari}},\ }\bibfield  {title} {\bibinfo {title} {Anisotropic perturbations
  in three-dimensional o($n$)-symmetric vector models},\ }\href
  {https://doi.org/10.1103/PhysRevB.84.125136} {\bibfield  {journal} {\bibinfo
  {journal} {Phys. Rev. B}\ }\textbf {\bibinfo {volume} {84}},\ \bibinfo
  {pages} {125136} (\bibinfo {year} {2011})}\BibitemShut {NoStop}%
\bibitem [{\citenamefont {Hasenbusch}(2001)}]{o3-MC2}%
  \BibitemOpen
  \bibfield  {author} {\bibinfo {author} {\bibfnamefont {M.}~\bibnamefont
  {Hasenbusch}},\ }\bibfield  {title} {\bibinfo {title} {Eliminating leading
  corrections to scaling in the three-dimensional o(n)-symmetric $\phi^4$
  model: N= 3 and 4},\ }\href {https://doi.org/10.1088/0305-4470/34/40/302}
  {\bibfield  {journal} {\bibinfo  {journal} {Journal of Physics A:
  Mathematical and General}\ }\textbf {\bibinfo {volume} {34}},\ \bibinfo
  {pages} {8221} (\bibinfo {year} {2001})}\BibitemShut {NoStop}%
\bibitem [{\citenamefont {Kos}\ \emph {et~al.}(2016)\citenamefont {Kos},
  \citenamefont {Poland}, \citenamefont {Simmons-Duffin},\ and\ \citenamefont
  {Vichi}}]{o3-CB1}%
  \BibitemOpen
  \bibfield  {author} {\bibinfo {author} {\bibfnamefont {F.}~\bibnamefont
  {Kos}}, \bibinfo {author} {\bibfnamefont {D.}~\bibnamefont {Poland}},
  \bibinfo {author} {\bibfnamefont {D.}~\bibnamefont {Simmons-Duffin}},\ and\
  \bibinfo {author} {\bibfnamefont {A.}~\bibnamefont {Vichi}},\ }\bibfield
  {title} {\bibinfo {title} {Precision islands in the ising and o(n ) models},\
  }\href {https://doi.org/10.1007/JHEP08(2016)036} {\bibfield  {journal}
  {\bibinfo  {journal} {Journal of High Energy Physics}\ }\textbf {\bibinfo
  {volume} {2016}},\ \bibinfo {pages} {36} (\bibinfo {year}
  {2016})}\BibitemShut {NoStop}%
\bibitem [{\citenamefont {Echeverri}\ \emph {et~al.}(2016)\citenamefont
  {Echeverri}, \citenamefont {von Harling},\ and\ \citenamefont
  {Serone}}]{o3-CB2}%
  \BibitemOpen
  \bibfield  {author} {\bibinfo {author} {\bibfnamefont {A.~C.}\ \bibnamefont
  {Echeverri}}, \bibinfo {author} {\bibfnamefont {B.}~\bibnamefont {von
  Harling}},\ and\ \bibinfo {author} {\bibfnamefont {M.}~\bibnamefont
  {Serone}},\ }\bibfield  {title} {\bibinfo {title} {The effective bootstrap},\
  }\href {https://doi.org/10.1007/JHEP09(2016)097} {\bibfield  {journal}
  {\bibinfo  {journal} {Journal of High Energy Physics}\ }\textbf {\bibinfo
  {volume} {2016}},\ \bibinfo {pages} {97} (\bibinfo {year}
  {2016})}\BibitemShut {NoStop}%
\bibitem [{\citenamefont {Deng}(2006)}]{o4-MC1}%
  \BibitemOpen
  \bibfield  {author} {\bibinfo {author} {\bibfnamefont {Y.}~\bibnamefont
  {Deng}},\ }\bibfield  {title} {\bibinfo {title} {Bulk and surface phase
  transitions in the three-dimensional $o(4)$ spin model},\ }\href
  {https://doi.org/10.1103/PhysRevE.73.056116} {\bibfield  {journal} {\bibinfo
  {journal} {Phys. Rev. E}\ }\textbf {\bibinfo {volume} {73}},\ \bibinfo
  {pages} {056116} (\bibinfo {year} {2006})}\BibitemShut {NoStop}%
\bibitem [{\citenamefont {Kos}\ \emph {et~al.}(2015)\citenamefont {Kos},
  \citenamefont {Poland}, \citenamefont {Simmons-Duffin},\ and\ \citenamefont
  {Vichi}}]{o4-CB1}%
  \BibitemOpen
  \bibfield  {author} {\bibinfo {author} {\bibfnamefont {F.}~\bibnamefont
  {Kos}}, \bibinfo {author} {\bibfnamefont {D.}~\bibnamefont {Poland}},
  \bibinfo {author} {\bibfnamefont {D.}~\bibnamefont {Simmons-Duffin}},\ and\
  \bibinfo {author} {\bibfnamefont {A.}~\bibnamefont {Vichi}},\ }\bibfield
  {title} {\bibinfo {title} {Bootstrapping the o(n) archipelago},\ }\href
  {https://doi.org/10.1007/JHEP11(2015)106} {\bibfield  {journal} {\bibinfo
  {journal} {Journal of High Energy Physics}\ }\textbf {\bibinfo {volume}
  {2015}},\ \bibinfo {pages} {106} (\bibinfo {year} {2015})}\BibitemShut
  {NoStop}%
\bibitem [{\citenamefont {Okabe}\ and\ \citenamefont {Oku}(1978)}]{large-N}%
  \BibitemOpen
  \bibfield  {author} {\bibinfo {author} {\bibfnamefont {Y.}~\bibnamefont
  {Okabe}}\ and\ \bibinfo {author} {\bibfnamefont {M.}~\bibnamefont {Oku}},\
  }\bibfield  {title} {\bibinfo {title} {{$1/n$ Expansion Up to Order $1/n^2$.
  III: Critical Exponents $\gamma$ and $\nu$ for d=3}},\ }\href
  {https://doi.org/10.1143/PTP.60.1287} {\bibfield  {journal} {\bibinfo
  {journal} {Progress of Theoretical Physics}\ }\textbf {\bibinfo {volume}
  {60}},\ \bibinfo {pages} {1287} (\bibinfo {year} {1978})},\ \Eprint
  {https://arxiv.org/abs/https://academic.oup.com/ptp/article-pdf/60/5/1287/5192266/60-5-1287.pdf}
  {https://academic.oup.com/ptp/article-pdf/60/5/1287/5192266/60-5-1287.pdf}
  \BibitemShut {NoStop}%
\bibitem [{\citenamefont {Vasil'ev}\ \emph {et~al.}(1982)\citenamefont
  {Vasil'ev}, \citenamefont {Pis'mak},\ and\ \citenamefont
  {Khonkonen}}]{o10-largeN1}%
  \BibitemOpen
  \bibfield  {author} {\bibinfo {author} {\bibfnamefont {A.~N.}\ \bibnamefont
  {Vasil'ev}}, \bibinfo {author} {\bibfnamefont {Y.~M.}\ \bibnamefont
  {Pis'mak}},\ and\ \bibinfo {author} {\bibfnamefont {Y.~R.}\ \bibnamefont
  {Khonkonen}},\ }\bibfield  {title} {\bibinfo {title} {1/n expansion:
  Calculation of the exponent $\nu$ in the order $1/n^3$ by the conformal
  bootstrap method},\ }\href {https://doi.org/10.1007/BF01015292} {\bibfield
  {journal} {\bibinfo  {journal} {Theoretical and Mathematical Physics}\
  }\textbf {\bibinfo {volume} {50}},\ \bibinfo {pages} {127} (\bibinfo {year}
  {1982})}\BibitemShut {NoStop}%
\bibitem [{\citenamefont {Broadhurst}\ \emph {et~al.}(1997)\citenamefont
  {Broadhurst}, \citenamefont {Gracey},\ and\ \citenamefont
  {Kreimer}}]{o10-largeN2}%
  \BibitemOpen
  \bibfield  {author} {\bibinfo {author} {\bibfnamefont {D.~J.}\ \bibnamefont
  {Broadhurst}}, \bibinfo {author} {\bibfnamefont {J.~A.}\ \bibnamefont
  {Gracey}},\ and\ \bibinfo {author} {\bibfnamefont {D.}~\bibnamefont
  {Kreimer}},\ }\bibfield  {title} {\bibinfo {title} {Beyond the triangle and
  uniqueness relations: non-zeta counterterms at large n from positive knots},\
  }\href {https://doi.org/10.1007/s002880050500} {\bibfield  {journal}
  {\bibinfo  {journal} {Zeitschrift f{\"u}r Physik C Particles and Fields}\
  }\textbf {\bibinfo {volume} {75}},\ \bibinfo {pages} {559} (\bibinfo {year}
  {1997})}\BibitemShut {NoStop}%
\bibitem [{\citenamefont {Stanley}(1968)}]{on-nvect}%
  \BibitemOpen
  \bibfield  {author} {\bibinfo {author} {\bibfnamefont {H.~E.}\ \bibnamefont
  {Stanley}},\ }\bibfield  {title} {\bibinfo {title} {Dependence of critical
  properties on dimensionality of spins},\ }\href
  {https://doi.org/10.1103/PhysRevLett.20.589} {\bibfield  {journal} {\bibinfo
  {journal} {Phys. Rev. Lett.}\ }\textbf {\bibinfo {volume} {20}},\ \bibinfo
  {pages} {589} (\bibinfo {year} {1968})}\BibitemShut {NoStop}%
\bibitem [{\citenamefont {{de Gennes}}(1972)}]{o0-first}%
  \BibitemOpen
  \bibfield  {author} {\bibinfo {author} {\bibfnamefont {P.}~\bibnamefont {{de
  Gennes}}},\ }\bibfield  {title} {\bibinfo {title} {Exponents for the excluded
  volume problem as derived by the wilson method},\ }\href
  {https://doi.org/https://doi.org/10.1016/0375-9601(72)90149-1} {\bibfield
  {journal} {\bibinfo  {journal} {Physics Letters A}\ }\textbf {\bibinfo
  {volume} {38}},\ \bibinfo {pages} {339 } (\bibinfo {year}
  {1972})}\BibitemShut {NoStop}%
\bibitem [{\citenamefont {Gaspari}\ and\ \citenamefont
  {Rudnick}(1986)}]{o0-second}%
  \BibitemOpen
  \bibfield  {author} {\bibinfo {author} {\bibfnamefont {G.}~\bibnamefont
  {Gaspari}}\ and\ \bibinfo {author} {\bibfnamefont {J.}~\bibnamefont
  {Rudnick}},\ }\bibfield  {title} {\bibinfo {title} {n-vector model in the
  limit n\ensuremath{\rightarrow}0 and the statistics of linear polymer
  systems: A ginzburg-landau theory},\ }\href
  {https://doi.org/10.1103/PhysRevB.33.3295} {\bibfield  {journal} {\bibinfo
  {journal} {Phys. Rev. B}\ }\textbf {\bibinfo {volume} {33}},\ \bibinfo
  {pages} {3295} (\bibinfo {year} {1986})}\BibitemShut {NoStop}%
\bibitem [{Note1()}]{Note1}%
  \BibitemOpen
  \bibinfo {note} {This would be the case if we attempted to use (\ref
  {eq:regulator}) at N${}^3$LO of the DE.}\BibitemShut {Stop}%
\end{thebibliography}%

\appendix
\section{The theta-regulator as a limit of a continuous regulator}\label{app:theta-reg}
\begin{figure}[hbt!]
\includegraphics[width=0.9\linewidth,height=50mm]{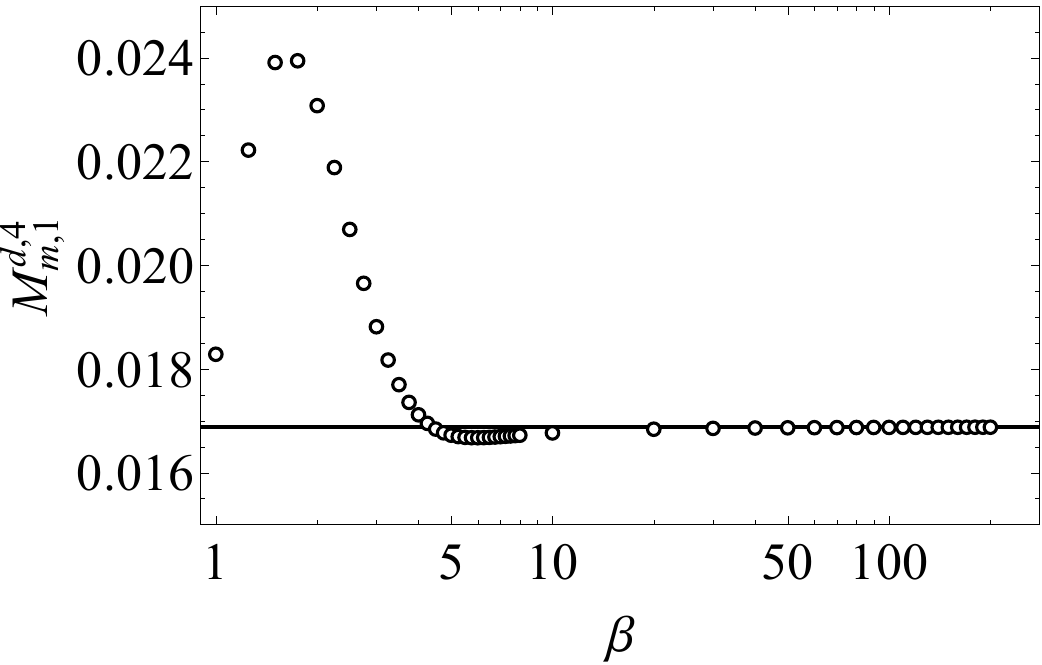}
\caption{\label{fig:integral-proof}The threshold function $M_{m,1}^{d,4}$ evaluated by numerical integration of the integral (\ref{eq:app-fig-eq}) (open circles) at different values of $\beta$ versus the analytical result (corresponding to $\beta\to\infty$) shown in Eq.~(\ref{eq:ambiguous}) (straight line). We considered dimensionless variables and set $d=3$, $\alpha=\tilde{\omega}=\tilde{Z}=\tilde{W}=1$ for the sake of example. 
}
\end{figure}

Regulators, which are not $C^\infty$ functions are not applicable beyond a certain order in the DE. Some threshold integrals at the NNLO of the DE evaluated with (\ref{eq:regulator}) are ambiguous, or even undefined \footnote{This would be the case if we attempted to use (\ref{eq:regulator}) at N${}^3$LO of the DE.} when $\delta(0)$ appears after performing the integration of the threshold functions.  The purpose of this appendix is to prove that the ambiguity of the threshold integrals is lifted when one considers (\ref{eq:regulator}) as the limit of a $C^\infty$-type regulator. We consider
\begin{equation}\label{eq:betareg}
    r_{\beta}(y)= \alpha \frac{(1-y)^2}{y}\frac{1}{1+e^{-2\beta(1-y)}},
\end{equation}
with the property
\begin{equation}
    \lim_{\beta\to\infty}r_{\beta}(y) = r_\Theta(y).
\end{equation}
Actually, the only ambiguous integral with (\ref{eq:regulator}) is of type $M_{m,1}^{d,4}$. Thus we are going to compute this integral with (\ref{eq:betareg}) and prove that in the $\beta\to\infty$ limit we unambiguously recover the result in Eq.~(\ref{eq:ambiguous}). It is convenient to introduce the integration variable $\epsilon = y-1$ %so that $y=1+\epsilon$ 
and to compute explicitly the derivatives of (\ref{eq:betareg}), resulting in
\begin{equation}\label{eq:app-fig-eq}
    M_{m,1}^{d,4} = 
    -128\alpha^2(Z_k^2 k^{d+2})\frac{\Omega_d}{(2\pi)^d} I_\beta
\end{equation}
where
\begin{equation}\label{eq:app-fig-eq}
\bsp
    I_\beta &= \int_{-1}^\infty\!\rd\epsilon  \frac{(1+\epsilon)^{-1+d/2}}{(\omega+(1+\epsilon)Z+(1+\epsilon)^2 W + \alpha \frac{\epsilon^2}{1+e^{2\beta\epsilon})})^m}
 \\&\qquad\times   
\epsilon \times \mathcal{I}.
\esp
\end{equation}
Here we organized all the derivatives of the regulator into the function $\mathcal{I}$:
\begin{equation}
\bsp
    \mathcal{I} &= \beta^2\frac{e^{6\beta\epsilon}}{(1+e^{2\beta\epsilon})^7}
    \biggl(\beta  \epsilon  (\epsilon +1) \sinh (\beta  \epsilon )+
 \\&+     
    (\epsilon  (\beta  \epsilon +\beta -1)-2) \cosh (\beta  \epsilon ))\biggr) \biggl[(3 (\sinh (\beta  \epsilon) 
 \\&+     
    \sinh (3 \beta  \epsilon ))+\beta  \epsilon  (\beta  \epsilon  (\sinh (3 \beta  \epsilon )-11 \sinh
   (\beta  \epsilon ))
 \\&+     
   12 \cosh (\beta  \epsilon )-4 \cosh (3 \beta  \epsilon )) \biggr]
   \,.
 \esp
\end{equation}
We have also written the inverse propagator $G$ explicitly and $\omega$, $Z$ and $W$ correspond to the scale dependent functions $2\rho U_k''(\rho)$, $Z_k(\rho)$ and $W_k(\rho)$.
If one considers the integral $I_\beta$ as the sum of three integration regions
\begin{equation}
    \int_{-1}^{-a}+\int_{-a}^{a}+\int_{a}^\infty
\end{equation}
with $0<a\ll 1$, then in the limit $\beta\to\infty$, the integrands of the integrals over the regions $[-1,-a]$ and $[a,\infty)$ vanish. Hence, in our computations we need the limit 
\begin{equation}
\lim_{\beta\to \infty} I_\beta =\lim_{\beta\to \infty} I_\beta(a)
\end{equation}
where
\begin{equation}\label{eq:app-integral}
\bsp
    I_\beta(a) &= \int_{-a}^a\!\rd\epsilon  \frac{(1+\epsilon)^{-1+d/2}}{(\omega+(1+\epsilon)Z+(1+\epsilon)^2 W + \alpha \frac{\epsilon^2}{1+e^{2\beta\epsilon})})^m}
 \\&\qquad\times   
\epsilon \times \mathcal{I}\,,
\esp
\end{equation}
with $a$ being a small positive integer, so we can expand the dimension dependent and inverse propagator part of $I_\beta(a)$ in Taylor series. At leading order (LO), this Taylor expansion leads to
\begin{equation}
\bsp
I^{\rm LO}_\beta(a) &= \frac{1}{(\omega+Z+ W)^m}\int_{-a}^a\!\rd\epsilon  
~\epsilon \times \mathcal{I}
\\&\equiv
\frac{1}{G(1)^m}\int_{-a}^a\!\rd\epsilon  
~\epsilon \times \mathcal{I},
\esp
\end{equation}
This integral can be computed analytically  and results in a very long combination of polynomials of $\beta$, $\epsilon$ and poly-logarithms, we do not show the explicit result here as it can be verified with the integrator of \texttt{Mathematica} for instance. Now, we are in the position to take the limit $\beta\to\infty$ of the integral, which is independent of $a$,
\begin{equation}\label{eq:app-fin}
    I^{\rm LO}_{\beta\to\infty} = -\frac{1}{32}\frac{1}{(\omega+Z+ W)^m}.
\end{equation}
One can of course take into account higher order terms in the Taylor expansion of the dimension dependent and inverse propagator part of (\ref{eq:app-integral}) so that the $n$-th order term in this expansion will be proportional to 
\begin{equation}
   \int_{-a}^a\!\rd\epsilon^{1+n}  
~\epsilon \times \mathcal{I}.
\end{equation}
Such higher order terms vanish in the limit $\beta\to\infty$, which we show here for the next-to-leading order (NLO) approximation -- also independent of $a$ -- to (\ref{eq:app-integral}),
\begin{equation}
\bsp
 I^{\rm NLO}_{\beta} &= -\frac{1}{32}\frac{1}{G(1)^m}\biggl[1+ \left(\frac{1}{\beta}\frac{7\pi^4-360}{1200}\right)
 \times
  \\&\times
 \biggl(\frac{d}{2}-1-m\frac{1+2W}{G(1)}\biggr)+\mathcal{O}\left(\frac{1}{\beta^3}\right)\biggr].
 \esp
\end{equation}
As the final result in the $\beta\to\infty$ limit we obtain 
\begin{equation}
    M_{m,1}^{d,4} = 
    4\alpha^2(Z_k^2 k^{d+2})\frac{\Omega_d}{(2\pi)^d} \frac{1}{G(1)^m},
\end{equation}
which coincides with (\ref{eq:ambiguous}). We conclude that the regulator is unambiguous at the NNLO of the DE once considered as the limit of a $C^\infty$ type regulator. A numerical example is also shown in Fig.~\ref{fig:integral-proof}.
We have claimed in the main text, that with the regulator (\ref{eq:regulator})
\begin{equation}
    M_{m,1}^{d,3} =0,
\end{equation}
due to the properties of the Dirac delta. Here we show that this integral with the regulator (\ref{eq:betareg}) indeed vanishes in the limit $\beta\to \infty$. Using an identical derivation as used for $M_{m,1}^{d,4}$ above, it is straightforward to show that
\begin{equation}\label{eq:app-md3}
\bsp
 M_{m,1}^{d,3} &= \lim_{\beta\to \infty}
 \left(-\alpha^2 (Z_k^2 k^{d+2})\frac{\Omega_d}{(2\pi)^d}\frac{1}{G(1)^m}\right)
 \\&\qquad\quad\times
 \biggl[\biggl(\frac{7\pi^4-360}{450\beta}\biggr)+\mathcal{O}\biggl(\frac{1}{\beta^3}\biggr)\biggr].
 \esp
\end{equation}
As check, one can compute numerically the integral $M_{m,1}^{d,3}$ for arbitrary values of $\beta$ using the regulator (\ref{eq:betareg}) and compare it to the analytical result in Eq.~(\ref{eq:app-md3}). Using the numerical integrations similar to those used for Fig.~\ref{fig:integral-proof}, we obtain $-1.5100\times10^{-5}$ from the direct numerical integration and $-1.5098\times10^{-5}$ from the analytical result (\ref{eq:app-md3})  at $\beta=200$. The integrands of $M_{m,1}^{d,2}$ and $M_{m,1}^{d,1}$ using the regulator (\ref{eq:betareg}) unambiguously reduce to those corresponding to the $\Theta$-regulator (\ref{eq:regulator}) in the limit $\beta\to\infty$. This is also the case for the threshold integrals $N_{m,b,c}^{d+a,\beta,\gamma}$, when $\beta$ and $\gamma$ are $1$ or $2$. If either $\beta$ or $\gamma$ is $3$, then the integral behaves as $M_{m,1}^{d,3}$, which we have already discussed.

\section{The error estimates and central values}\label{app:error-estimate}
In this work we follow the instructions of Ref.~\cite{full-on} for appropriate error bars. However, due to the polynomial expansion an additional source of error appears. We summarize here the steps we take in this work to obtain the final prediction for the exponent $X$ and also to obtain its uncertainty. First, at a given order ($\partial^s$) of the DE we compute the PMS optimized value for various order $M$ of the polynomial truncation of the scale dependent functions (where $M$ belongs to the least truncated case) for the regulators (\ref{eq:regulator}) and (\ref{eq:exp-regulator}). This way we obtain the set of raw data $\{X^{(s),opt}_{M,\Theta},X^{(s),opt}_{M,exp}\}$. Let us now discuss the computation of the final values and the different sources of uncertainties considered in this work point by point.

(i) One can choose for the final result $\overline{X}^{(s)}$ at a given order of the DE the central value $\overline{X}^{(s)}\equiv \overline{X}^{(s)}_M =(X^{(s),opt}_{M,\Theta}+X^{(s),opt}_{M,exp})/2$, which is indeed our choice at LPA and NLO approximations. However we apply further improvement to the NNLO result. Namely, we first apply the ansatz (\ref{eq:resumtaylor}) on the NNLO dataset to extrapolate to the exponents corresponding to $M_{WHJ}\to\infty$ and after this step we compute the central value of the result from the regulators. We consider the values obtained this way to be our NNLO prediction, with the polynomial truncation improvement. Furthermore, we extrapolate our predictions at NNLO employing Wynn's epsilon algorithm whenever it is applicable, i.e.~when the predictions at successive orders of the DE show an alternating behaviour. We cited those extrapolated predictions in Tab.~\ref{tab:on-refsum}.

(ii) One source of uncertainty originates from the choice of regulators, which we denote by $\Delta_{reg}\overline{X}^{(s)}$ corresponding to the prediction $\overline{X}^{(s)}$. We define it to be half of the largest difference $\Delta_{reg}\overline{X}^{(s)}_{M}=|X^{(s),opt}_{M,\Theta}-X^{(s),opt}_{M,exp}|/2$ between the two predictions obtained with different regulators. Considering empirical data, such as in Ref.~\cite{full-on} we see that the $\Theta_n$-type regulator, with the smallest possible $n$ at the given order in the DE yields predictions closest to the most precise ones, obtained from other methods. The exponential regulator (\ref{eq:exp-regulator}) on the other hand seems to produce predictions farthest from the most precise ones. This is true at least up to NNLO, which supports our choice for $\Delta_{reg}\overline{X}^{(s)}$ at least up to NNLO of the DE.

(iii) Next, we compute the uncertainty of the DE according to Ref.~\cite{full-on}, exploiting the hidden small parameter $1/4-1/9$ of the DE. Calling this source of error $\Delta_{DE}\overline{X}^{(s)}$, we have $\Delta_{DE}\overline{X}^{(s)} = |\overline{X}^{(s)}-\overline{X}^{(s-2)}|/4$, where $\overline{X}^{(s-2)}$ corresponds to the result from the previous order ($\partial^{s-2}$) of the DE. Of course, this implies that we are unable to estimate the error from the DE at the LPA this way.

(iv) In addition to $\Delta_{reg}\overline{X}^{(s)}$ and $\Delta_{DE}\overline{X}^{(s)}$, the finite truncation $M$ of the scale dependent functions also introduces another source of systematic uncertainty $\Delta_{poly}\overline{X}^{(s)}$. Every order of the DE introduces new scale dependent functions $F$ and thus additional sources of uncertainty if we truncate them. Thus we have $\Delta_{poly}\overline{X}^{(s)}=\sum_{\{F\}} \Delta_{F}\overline{X}^{(s)}$ where we sum over all scale dependent functions $F$ available at the $\partial^{s}$-order of the DE. We define these independent contributions as $\Delta_{F}\overline{X}^{(s)} = |\overline{X}^{(s)}_{M_F}-\overline{X}^{(s)}_{M_F-1}|$. We do this because we go so far in the polynomial expansion that $\Delta_{F}\overline{X}^{(s)}$ decreases monotonically for higher values of $M_F$. To estimate $\Delta_{poly}\overline{X}^{(4)}$ at NNLO, we take the absolute difference between the data improved by (\ref{eq:resumtaylor}). We have already elaborated in the main text, that we apply the same degree of truncation to all scale dependent functions corresponding to a given order of the DE. For instance, in the $Z_2$ symmetric models at NNLO, we have $\Delta_{poly}\overline{X}^{(4)}=\Delta_{U}\overline{X}^{(4)}+\Delta_{Z}\overline{X}^{(4)}+\Delta_{WHJ}\overline{X}^{(4)}$. With the truncation used in this work, we have $\Delta_{U}\overline{X}^{(4)}<\Delta_{Z}\overline{X}^{(4)}\ll \Delta_{WHJ}\overline{X}^{(4)}$, so that $\Delta_{poly}\overline{X}^{(4)}\approx \Delta_{WHJ}\overline{X}^{(4)}$.

    (v) Finally we have to combine the different sources of uncertainties in order to obtain the total uncertainties $\Delta\overline{X}^{(s)}$ quoted in the tables of the main text. There is no straightforward way to prove that the discussed sources are uncorrelated, so we decided to use a simple sum,
\begin{equation}
    \Delta\overline{X}^{(s)} = \Delta_{DE}\overline{X}^{(s)}+\Delta_{reg}\overline{X}^{(s)}+\Delta_{poly}\overline{X}^{(s)}
\end{equation}
as conservative estimate.
As mentioned in point (ii) $\Delta_{DE}\overline{X}^{(0)}$ is unavailable. Furthermore, $\Delta_{reg}\overline{X}^{(0)}\gg \Delta_{poly}\overline{X}^{(0)}$. Thus in practice we have $\Delta\overline{X}^{(0)} = \Delta_{reg}\overline{X}^{(0)}$ at LPA. At the NLO $\Delta_{DE}\overline{X}^{(2)}$ is the dominant source of uncertainty.

\section{Technical details of numerical computations}
We locate the Wilson-Fisher fixed point corresponding to the complete set of beta-functions ($\{\beta_{\tilde{g}_i} =0\}$) for the dimensionless couplings $\tilde{g}_i$. In order to find the nontrivial root of this system of equations we have used the Affine Covariant Newton method with the iterative algorithm detailed in Sec.~\ref{sec:z2}. and \ref{sec:on-num}. In the rare case it did not converge in $100$ iterations we further applied the secant method. This requires two initial values, to obtain those we simply multiply the output from the Affine Covariant Newton method with $0.9$ and $1.1$. 

The numerical integration of the threshold integrals $L,M$ and $N$ (from Sec.~\ref{sec:threshint}) are computed with the optimized \texttt{NIntegrate} command of \texttt{Mathematica}, which selects the Gauss-Konrod quadrature formula as the most efficient numerical integration method.

In every instance we have worked with $12$ or more digits of precision in our numerical computations.

\section{Subleading scaling corrections}

We also provide the scaling corrections $\omega<\omega_1<\omega_2<\ldots$ to the correlation length as discussed in Sect.~\ref{sec:polyexp}. 
These smallest one $\omega$ is shown in Tab.~\ref{tab:on-data} for the $O(N)$ model at various $N$ values. 
The larger scaling corrections $\omega_1$, $\omega_2$ are summarized in Tab.~\ref{tab:on-sublead}. Generally $\omega_n$ becomes more susceptible  to the polynomial truncation with increasing $n$, $\omega_1$, $\omega_2$ are only stable in the first two or three significant digits with our employed truncation, detailed in Sec.~\ref{sec:on-num}. 
\begin{table}[hbt!]
    \centering
\begin{tabular}{ |c|c||c c |c| } 
\hline
$N$ & Order of DE & $\omega_1$ & $\omega_2$ \\
\hline
\hline
\multirow{3}{2cm}{~~~0} 
& LPA & - & $3.3$ \\ 
& NLO & $1.4$ & $4.0$ \\ 
& NNLO & $1.4$ & $3.3$ \\ 
\hline
\hline
\multirow{3}{4em}{1} 
& LPA & - & $3.2$ \\ 
& NLO & $1.7$ & $3.9$ \\ 
& NNLO & $1.7$ & $3.2$ \\ 
\hline
\hline
\multirow{3}{4em}{2} 
& LPA & - & $3.1$ \\ 
& NLO & $1.9\pm0.1i$ & $3.6$ \\ 
& NNLO & $1.8$ & $3.3$ \\ 
\hline
\hline
\multirow{3}{4em}{3} 
& LPA & - & $3.0$ \\ 
& NLO & $2.0\pm 0.5i$ & $3.5$ \\ 
& NNLO & $1.9$ & $3.4$ \\ 
\hline
\hline
\multirow{3}{4em}{4} 
& LPA & - & $2.94$ \\
& NLO & $1.9$ & $3.4$ \\ 
& NNLO & $1.9$ & $3.3$ \\ 
\hline
\hline
\multirow{3}{4em}{10} 
& LPA & - & $2.90$ \\
& NLO & $1.96$ & $2.8$ \\ 
& NNLO & $1.96$ & $2.9$ \\ 
\hline
\hline
\multirow{3}{4em}{100} 
& LPA & - & $2.99$ \\
& NLO & $2.00$ & $2.97$ \\ 
& NNLO & $1.99$ & $2.97$ \\ 
\hline
\end{tabular}
    \caption{The first two subleading scaling corrections $\omega_1$ and $\omega_2$  at the LPA, NLO and NNLO of the DE for the $O(N)$ symmetric models in $d=3$ Euclidean dimensions. We have only kept the first few significant digits, which 
     coincide for the predictions computed from the $\Theta$-regulator (\ref{eq:regulator}) and the exponential regulator (\ref{eq:exp-regulator}).}
    \label{tab:on-sublead}
\end{table}
\end{document}